\documentclass[usenatbib,usegraphicx]{mn2e}

\usepackage{aas_macros}
\usepackage{multirow}
\title[Statistics of Galactic Satellites]{
Spatial and luminosity distributions of galactic satellites}
\author[Guo et al]{Quan Guo, Shaun Cole, Vincent Eke, Carlos Frenk, John Helly\\
Institute for Computational Cosmology, Department of Physics, 
Durham University, Science
Laboratories, South Rd, Durham DH1 3LE.} 

\newcommand{\gf}{GALFORM~}
\newcommand{\satmc}{M_{\rm{c}}}
\newcommand{\satdmbin}{\Delta M_{\rm{bin}}}
\newcommand{\satdmf}{\Delta M_{\rm{faint}}}
\newcommand{\satdzs}{\Delta z_{\rm{s}}}

\newcommand{\ri}{R_{\rm{inner}}}
\newcommand{\ro}{R_{\rm{outer}}}

\newcommand{\msatt}{M_{\rm sat}^{\rm trun}}

\newcommand{\lcdm}{$\rm{\Lambda CDM}$~}

\newcommand\lsim{\mathrel{\hbox{\rlap{\hbox{\lower4pt\hbox{$\sim$}}}\hbox{$<$}}}}
\begin{document}


\maketitle

\label{firstpage}

\begin{abstract}
  We investigate the luminosity functions (LFs) and projected number
  density profiles of galactic satellites around isolated primaries of
  different luminosity. We measure these quantities for model
  satellites placed into the Millennium and Millennium II dark matter
  simulations by the \gf semi-analytic galaxy formation model for
  different bins of primary galaxy magnitude and we investigate their
  dependence on satellite luminosity.  We compare our model
  predictions to the data of Guo et al. from the Sloan Digital Sky
  Survey Data Release 8 (SDSS DR8). First, we use a mock light-cone
  catalogue to verify that the method we used to count satellites in
  the SDSS DR8 is unbiased.  We find that the radial distributions of
  model satellites are similar to those around comparable primary
  galaxies in the SDSS DR8, with only slight differences at low
  luminosities and small projected radii. However, when splitting the
  satellites by colour, the model and SDSS satellite systems no longer
  resemble one another, with many red model satellites, in contrast to
  the dominant blue fraction at similar luminosity in the SDSS. The few
  model blue satellites are also significantly less centrally
  concentrated in the halo of their stacked primary than their SDSS
  counterparts. The implications of this result for the \gf model are
  discussed.
\end{abstract}

\begin{keywords}
galaxies: dwarf, galaxies: structure, (galaxies:) Local Group, Galaxies:
fundamental parameters, galaxies: statistics

\end{keywords}

\section{Introduction}

While the standard $\Lambda$ Cold Dark Matter ($\rm\Lambda CDM$) model
has been shown to be in good agreement with observations of large
scale structure, the verification of this model at small, galactic
scales remains less certain. One reason for this is the increased
importance of astrophysical processes relative to gravity in this
strongly non-linear regime.  The study of the properties and
distribution of galactic satellite galaxies provides an opportunity to
test $\rm\Lambda CDM$ on small scales while also constraining
different aspects of galaxy formation models related to the rates at
which satellites form stars, become disrupted and merge with the
central galaxy.

The Local Group satellite system within the haloes of the Milky Way
(MW) and M31 is often the focus of studies of galactic satellites
because it is here where the lowest luminosity satellites can be
detected. While this system has been used in attempts to constrain the
cosmological model \citep[e.g][]{bull00,ben02,kly02,lov12,jie12a}, it is not
clear that it is typical of the population as a whole. Using the
technique of abundance matching in the Millennium II simulation
(MS-II), a large-volume, high-resolution dark matter simulation,
\cite{boy10} concluded that there should be significant scatter in the
properties of satellite systems from one primary to another. This
conclusion was supported by \cite{qi11} who applied a semi-analytic
galaxy formation model to the same simulation. Thus, a large,
statistically representative sample of primary galaxies is clearly
needed to test cosmological and galaxy formation models. Such an
approach also avoids the difficulty of having to define quite what is
meant by a Local Group.

The construction of large galaxy redshift surveys, such as the Two Degree Field
Galaxy Redshift Survey \citep[2dFGRS,][]{col01}, the SDSS \citep{yor00}, the
Galaxy and Mass Assembly \citep[GAMA,][]{dri09,dri11} Survey and the Deep
Extragalactic Evolutionary Probe 2 \citep[DEEP2,][]{dav03} Survey, has led to
the accumulation of external galaxy samples covering large volumes.  Many
studies of the luminosity function (LF), spatial distribution and kinematics of
bright satellites have been carried out using these and even earlier data sets
\citep[e.g.][]{zar93,zar97,van04,con05,van05,chen06,kop09,bus10,pre11}.  The
inclusion of a statistical background subtraction in the satellite system
estimators has allowed fainter satellites to be studied using deeper photometric
galaxy catalogues
\citep[e.g][]{lor94,liu08,liu10,lar11,my11,nie11,tal11,my12,jiang12,str12,tal12,wang12}.
By extending the regime over which the satellite distributions have been
quantified, a more stringent test of the models can be performed.

While some studies have attempted to make model predictions using
cosmological hydrodynamic simulations
\citep[e.g][]{lib07,oka09,oka10,wad11,par12}, these efforts are
limited to very few primary galaxies because of the high computational
cost.  The best way to make a statistical sample of model galaxies is
by combining large cosmological dark matter simulations, such as the
Millennium Simulation \citep[MS,][]{spr05} or the MS-II \citep{boy09},
with a method to include galaxies.  This approach was adopted by
\cite{van05}, who used the conditional luminosity function technique
that simultaneously optimizes the model match to the abundance and
clustering of low luminosity galaxies in order to study the satellite
projected number density profile.  \cite{chen06} investigated the same
statistic by assigning luminosities to dark matter structures so as 
to match the simulated cumulative circular velocity function to the
SDSS cumulative galaxy LF.  A similar subhalo abundance matching
method was used by \cite{busha11} to study the frequency of bright
satellites around MW-like primaries.

Semi-analytic models provide a more physically motivated approach to
including galaxies into dark matter simulations and have been shown to
match a wide variety of observational data
\citep[e.g][]{kau03,lac03,col94,kau99,som99,col00,ben02,bau05,bow06,cro06,qi11}.
A large number of studies have used this technique applied to
simulations such as the MS-II, Aquarius \citep{spr08} and others to
study various aspects of the galaxy population predicted in a
$\rm\Lambda CDM$ model
\citep[e.g][]{mun09,coo10,li10,mac10,fon11,jie12b,wang12}.  In particular, the
mock catalogues constructed by \cite{qi11} were tested against data from
the SDSS in two studies.  \cite{sal12} showed that the abundance of
satellite galaxies as a function of primary stellar mass in the SDSS
DR7 spectroscopic catalogue was in good agreement with this model.
Considering fainter dwarf satellites, \cite{wang12} studied the
luminosity, colour distribution and stellar mass function using SDSS
DR8 data, concluding that, apart from the model satellites becoming
red too quickly when entering the halo of the primary galaxy, many of
the observed trends were reproduced in the model.

In this paper, we will test the \lcdm model and the
semi-analytic galaxy formation model \gf \citep{bow06}, by comparing the
properties of model galactic satellite systems with those measured
from the SDSS DR8 spectroscopic and photometric galaxy catalogues.
We will introduce a new series of \gf model galaxy catalogues based on
MS-II that will allow us to extend the predictions from the MS to less
luminous galaxies.

In Section 2 we describe the SDSS and model galaxy catalogue data that
we use, and compare these two data sets.  Section 3 contains summaries
of the methods we use to select primaries and determine the satellite
luminosity function and the projected number density profile. These
two satellite galaxy distributions, upon which we will focus in this
paper, were calculated for the SDSS samples by Guo et al. (2011,
hereafter Paper I) and Guo et al. (2012, hereafter Paper II). A
verification that our estimators are unbiased is performed using the
model galaxy samples. The results of the comparison between model
satellite systems and those around similarly isolated SDSS primaries
is presented in Section~4. Implications for the model drawn from these
comparisons are discussed in Section~5; we conclude in Section~6.
Throughout the paper we assume a fiducial ${\rm\Lambda CDM}$
cosmological model with $\Omega_M=0.3$, $\Omega_{\Lambda}=0.7$ and
$H_0 = 70\ $km~s$^{-1}$Mpc$^{-1}$.

\section{Observed and model galaxies}

In this section we briefly review the data being used from the SDSS,
which were described in more detail in Paper~I and Paper~II, before
introducing the procedure to construct mock galaxy catalogues to
compare with these data.

\subsection{SDSS galaxies}

Galaxies from both the spectroscopic and photometric samples in the
SDSS DR8 are used for this study. Isolated primary galaxies, as
defined in Paper I and Paper II, are selected from
the spectroscopic survey, whereas satellites can come from either the
spectroscopic or photometric surveys. With the relatively poorly
constrained distances provided by the photometric redshifts, a
statistical background subtraction is performed to obtain an estimate
of the satellite galaxy population around each of the primaries, as
described in Section~\ref{sec:meth} below.

\subsection{Model galaxies}\label{ssec:model}

The model galaxy catalogues were created using a combination of large
dark matter simulations to define the mass distribution and a
semi-analytic model to place the galaxies within this density field.
Either the Millennium Simulation \citep[MS,][]{spr05} or the
Millennium-II Simulation \citep[MS-II,][]{boy09} was used to provide
the mass distribution. The former covers a large volume and hence
contains many suitable isolated primaries, while the latter traces the
mass distribution in a smaller volume, thus resolving structures
containing lower luminosity galaxies.  While the two simulations trace
the same number of particles ($\sim 10^{10}$), the MS and MS-II simulation
cubes are $714~{\rm Mpc}$ and $143~{\rm Mpc}$ long respectively. The
corresponding particle masses are $\sim1.23 \times 10^9~\rm M_{\sun}$
and $9.9\times10^6~\rm M_{\sun}$.

\begin{figure}
\includegraphics[width=80mm]{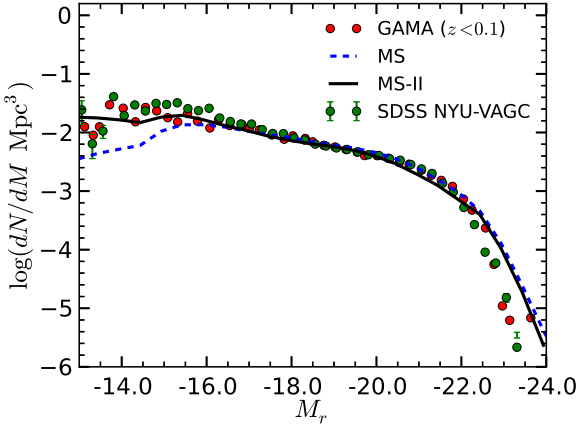}
\caption{The luminosity functions of all galaxies in 
     the MS (blue dashed line) and MS-II (black solid line) cubes
     compared with observed luminosity functions of galaxies
     in the GAMA survey (the $z<0.1$ subset from  Loveday et al. 2012, red
     points) and SDSS (Blanton et al. 2005, green points). 
     The observed luminosity functions are k-corrected (SDSS) or
     shifted (GAMA) to $z=0$.}
    \label{fig:lf}
\end{figure}

The model galaxies populate the dark matter structures according to
the galaxy formation model GALFORM \citep{bow06}, which includes
reionization at high redshift and energy injection from supernovae and
stellar winds in order to prevent the overproduction of low luminosity
galaxies. The luminosities of the most luminous galaxies are curbed by
feedback from AGN. Parameters in the treatment of the galaxy formation
processes have been chosen so as to produce as good a match as
possible to the observed K band LF of local
galaxies. Fig.~\ref{fig:lf} shows the luminosity functions of all
galaxies in the MS and MS-II simulation cubes. They match quite well
with both the observed $r$ band luminosity function of GAMA galaxies
\citep{gama12} shifted to $z=0$ by applying the offset
$r={}^{0.1}r-0.22$ \citep{bla05}, and the LF from the SDSS
\citep{bla05}. The drop in the MS LF at low luminosities reflects the
resolution limit, which corresponds roughly to $M_r=-16$. Galaxies
placed into the MS-II should be complete significantly beyond this and
thus suitable for studying faint satellites.

\begin{figure}
\centering
\includegraphics[width=80mm]{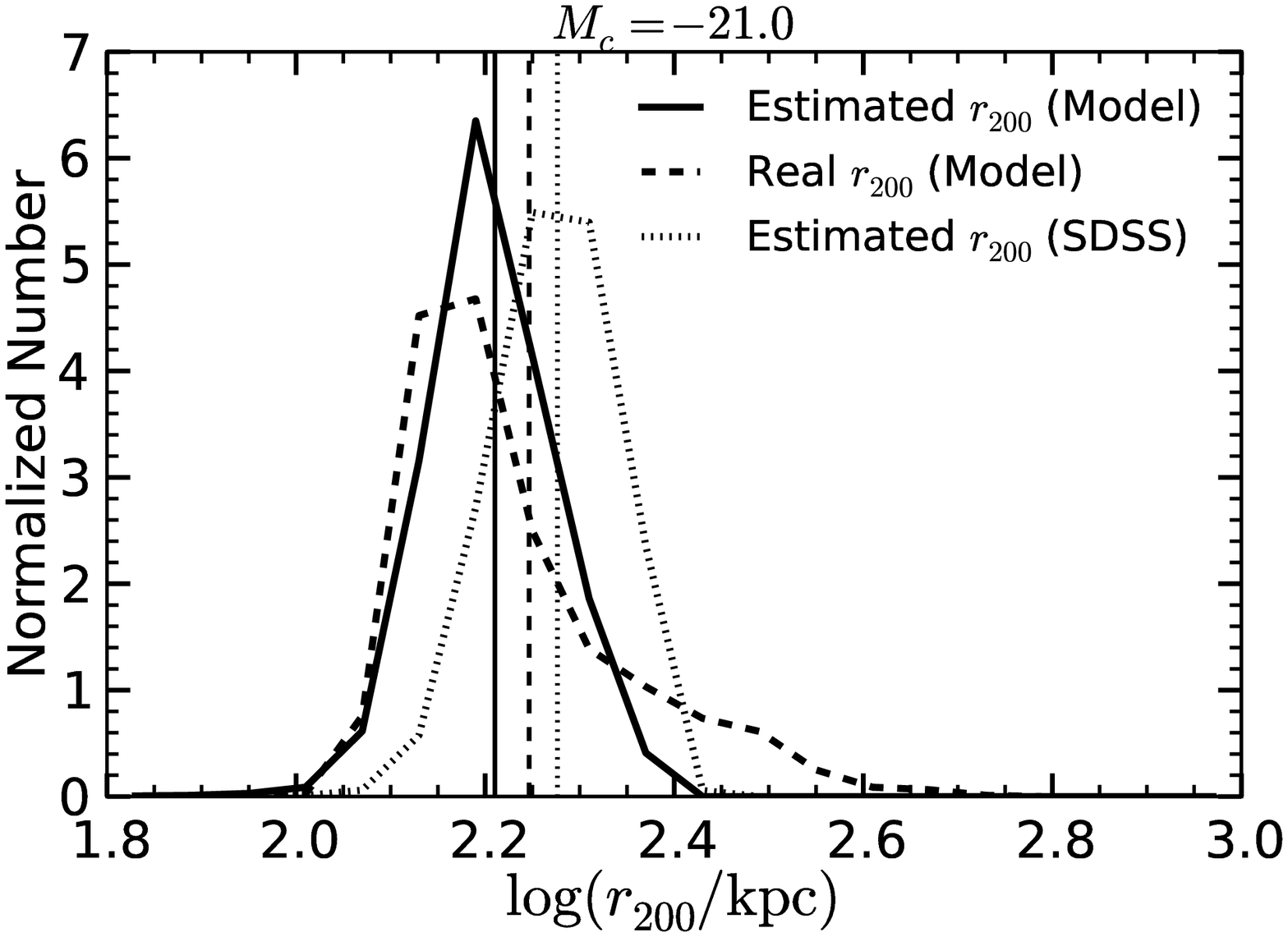}
\includegraphics[width=80mm]{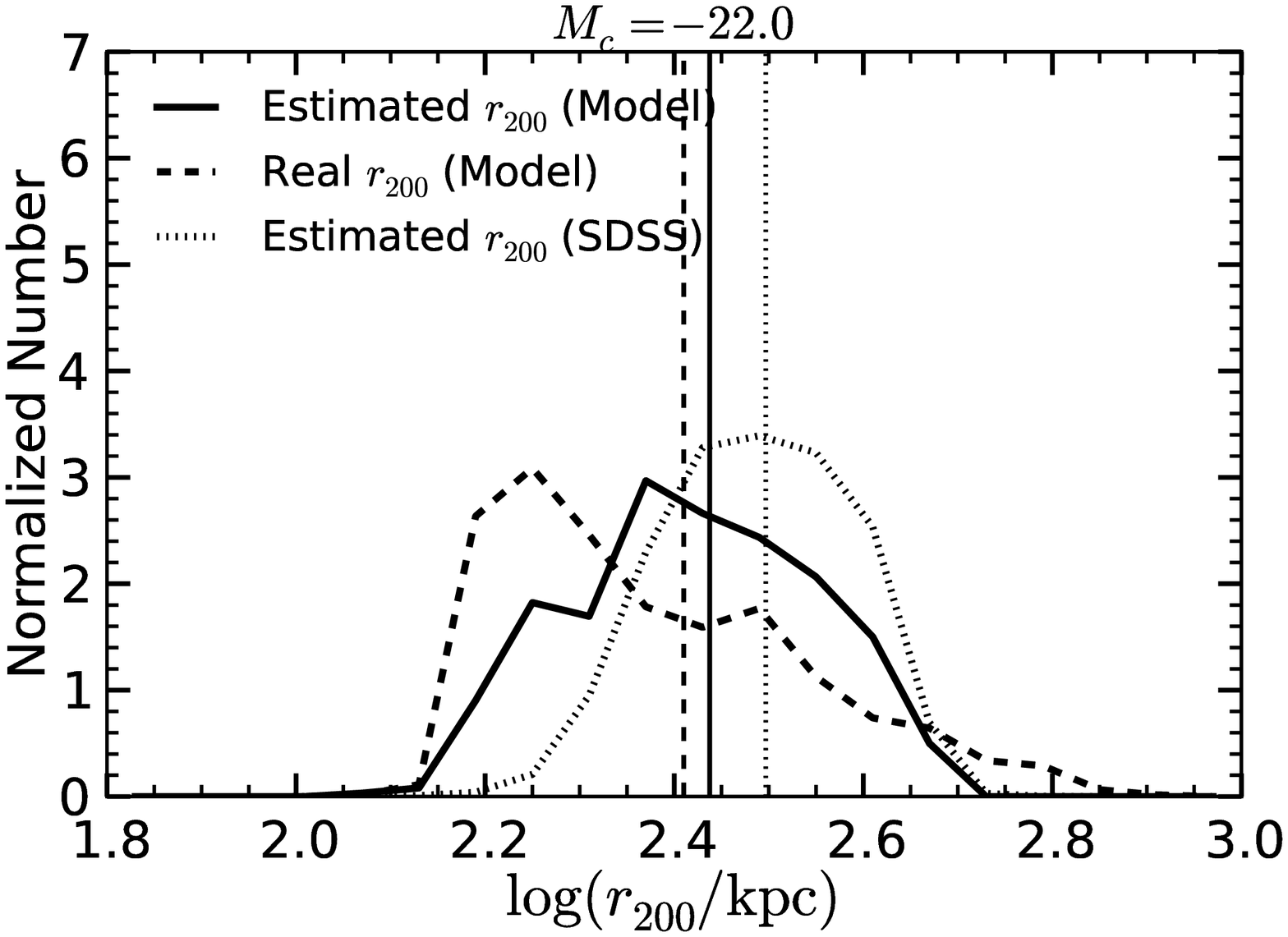}
\includegraphics[width=80mm]{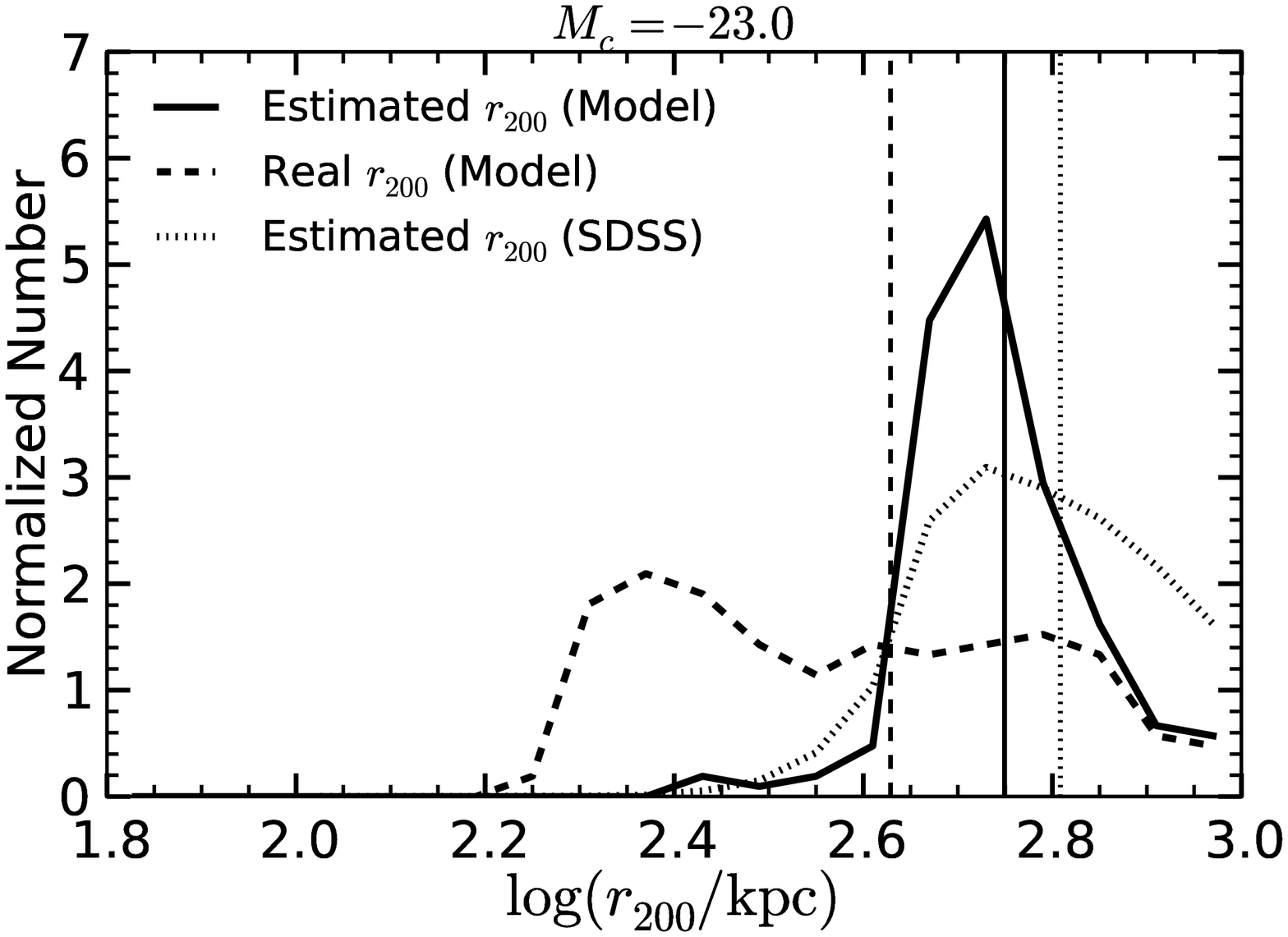}
\caption{The distribution of estimated $r_{200}$ values for model
  primary galaxies (solid lines), real $r_{200}$ values for their
  associated parent haloes (dashed lines) and estimated $r_{200}$
  values for primaries selected from the SDSS (dotted lines). From top
  to bottom, the panels show results for primary magnitude bins
  centred on $\satmc=-21.0, -22.0$ and $-23.0$ respectively. The
  vertical lines are the means of the corresponding distributions.}
\label{fig:fig_vr}
\end{figure}

\begin{figure*}
\includegraphics[width=80mm]{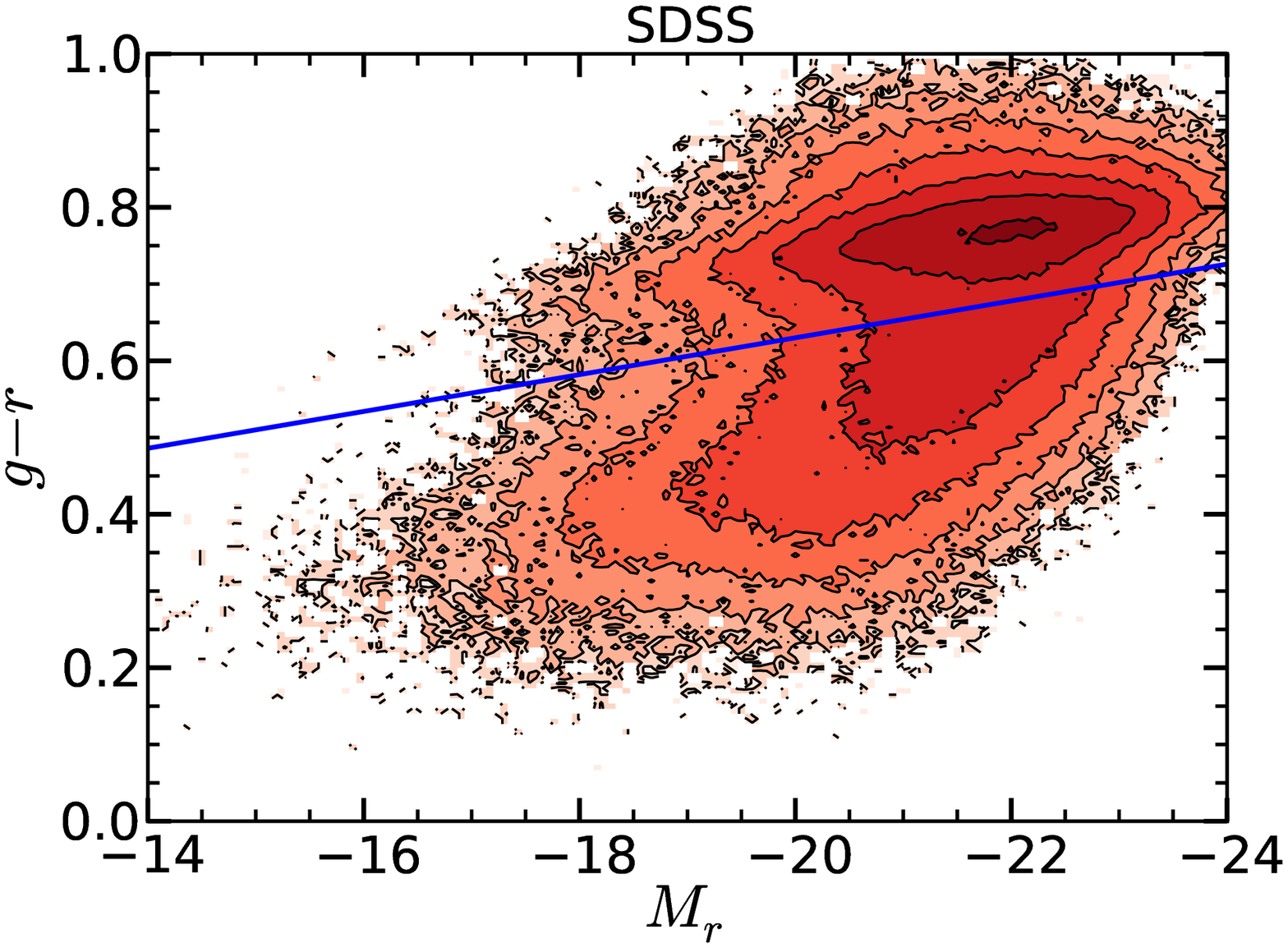}
\includegraphics[width=80mm]{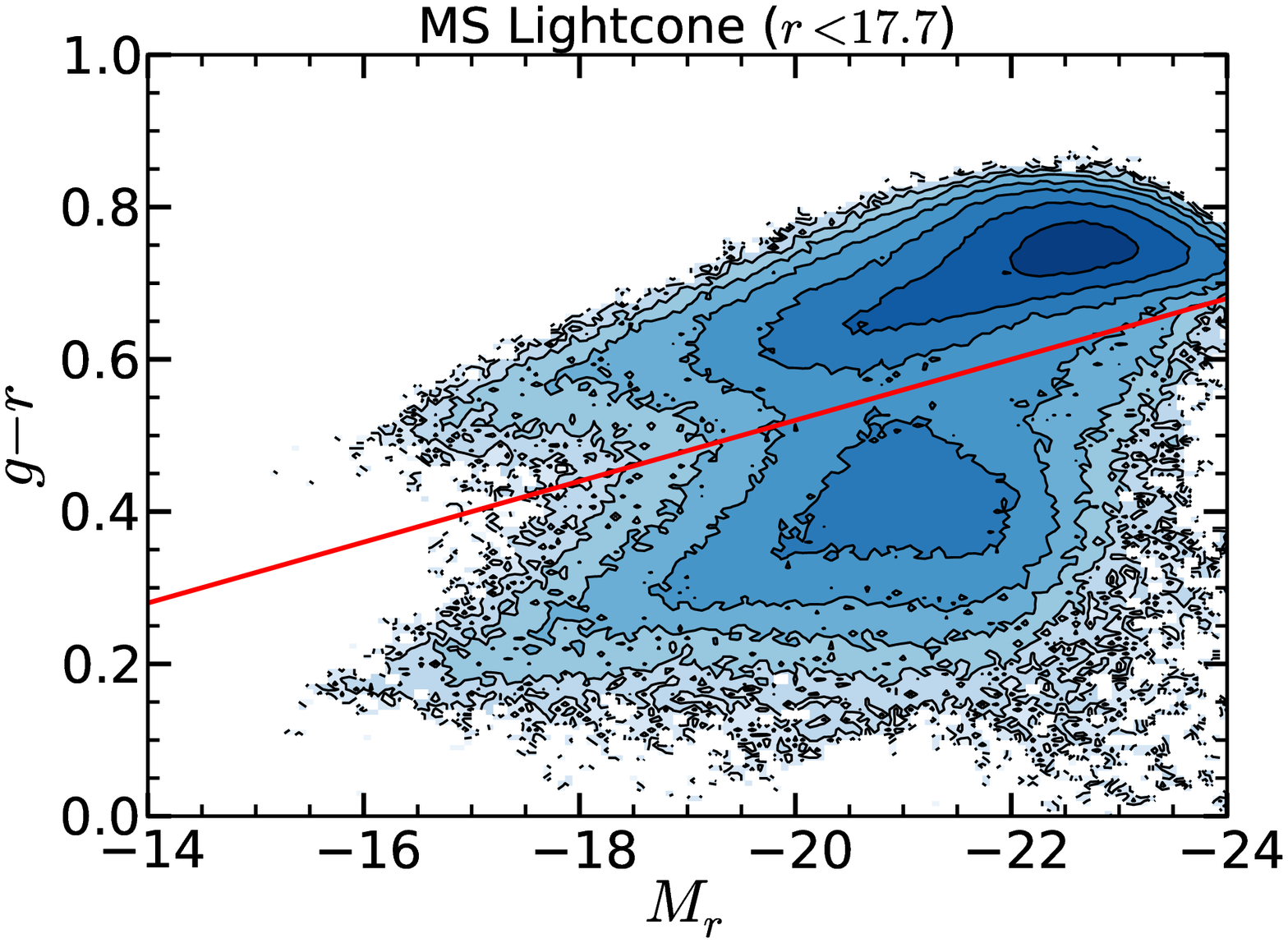}
\includegraphics[width=80mm]{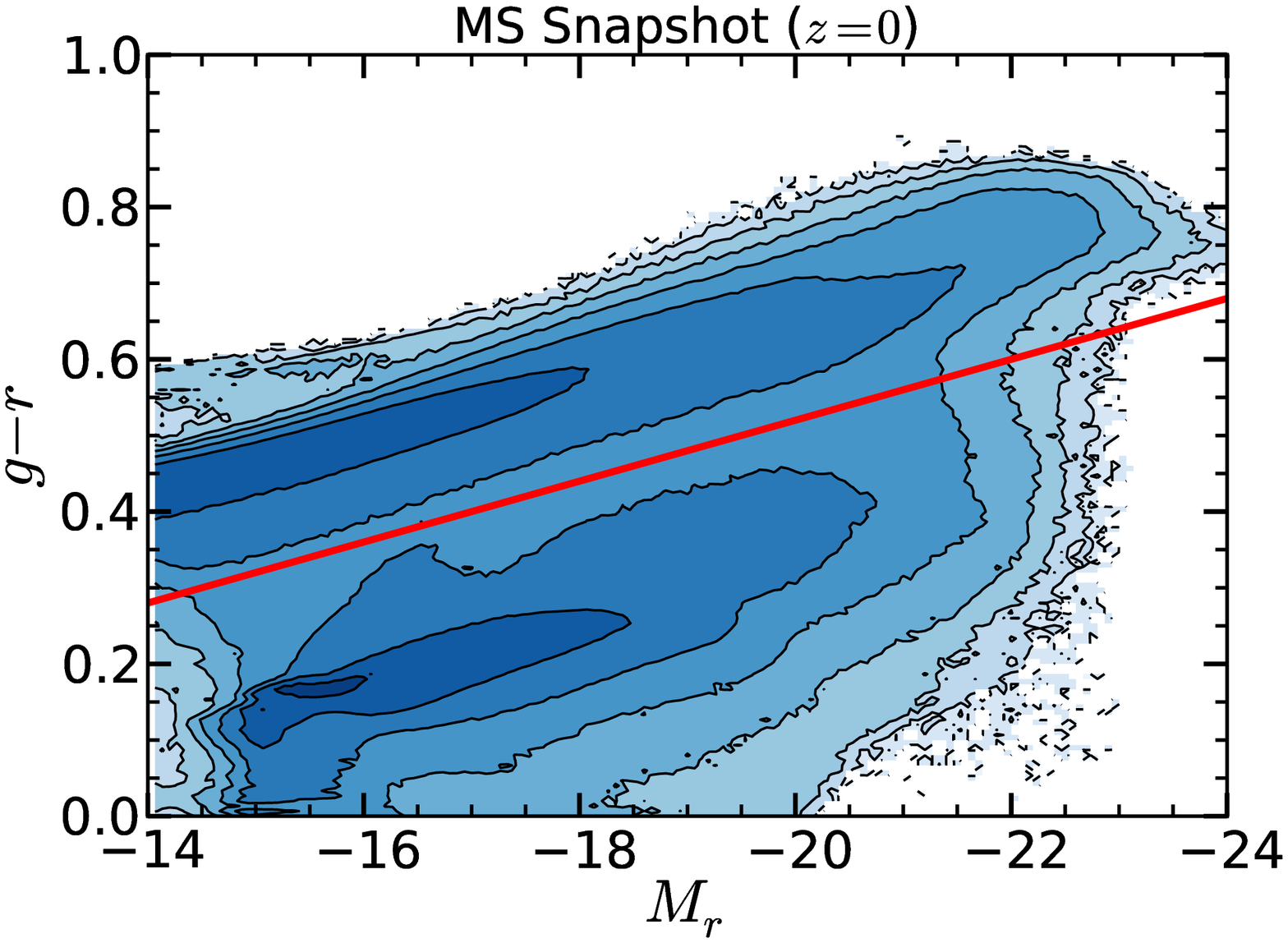}
\includegraphics[width=80mm]{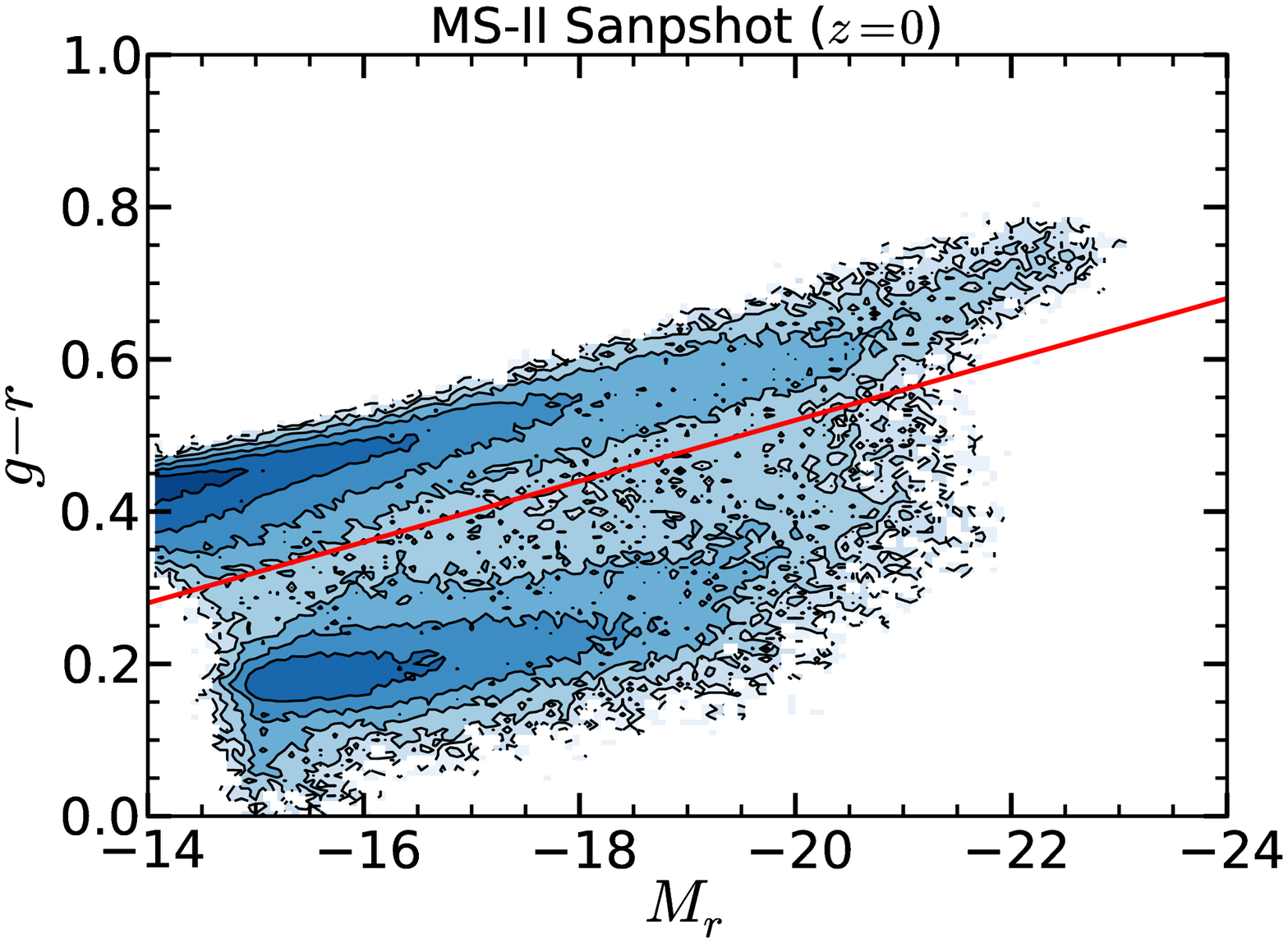}
\caption{Colour-magnitude diagrams of spectroscopic galaxies in SDSS
    (top left) and of galaxies in the MS mock light-cone (top right) and the two
    different simulation cubes (MS, bottom left and MS-II, bottom
    right). Contours represent lines of constant galaxy number density
    and the straight lines indicate the colour cuts 
    used to define red and blue primaries. The same cut is used for
    both the MS and MS-II.}
\label{fig:color}
\end{figure*}

For the MS, in addition to having the \gf model galaxies populating
the simulation cube, flux-limited light-cone mock catalogues, either with or
without the peculiar velocity included in the line-of-sight
`redshift', have also been constructed \citep{mer12}. These galaxies 
cover the redshift range $z=0.0-2.0$. To simulate the photometric
redshifts in the SDSS DR8 catalogue, we assign a photometric
redshift with a random error to every galaxy that is fainter than $m_r=17.7$.

Using both the MS and MS-II cubes of galaxies, we can test how robust
our results are to the numerical resolution of the dark matter
simulations.  Comparison of the results obtained from the MS cube with
those obtained from the light-cone mock catalogue, provides a test of
the accuracy of our background removal and satellite distribution
estimation procedures. Finally, the light-cone mock catalogue is
intended to mimic the SDSS survey and provide a direct test of the
model.

When calculating scaled satellite number density profiles, it is
necessary to determine the value of $r_{200}$ (defined as the radius
enclosing a mean total overdensity of 200 times the critical cosmic
value) associated with each primary galaxy. Following Paper II,
$r_{200}$ is estimated from the stellar mass, inferred from the
luminosity and colour of the primary. This is converted to a halo
mass, $M_{200}$, using the abundance matching technique of
\cite{guo10}, from which $r_{200}$ follows. The solid and dotted lines
in Fig.~\ref{fig:fig_vr} trace the distributions of $r_{200}$
estimated from the MS and SDSS primaries respectively, with the
different panels showing results for different luminosity
primaries. Vertical lines show the mean values of the distributions,
which differ by no more than about $15$ per cent in all cases. This
similarity between estimated satellite system sizes in the mock and
SDSS suggests that scaling the satellite number density profiles by
the system size should not create any large systematic differences
between the real and mock results.

The distributions of real $r_{200}$ values, inferred from the dark
matter distribution in the simulation cube, are shown with dashed
lines in Fig.~\ref{fig:fig_vr}.  For the lower luminosity primaries,
the real and estimated $r_{200}$ distributions are similar. However,
in the bottom panel of Fig.~\ref{fig:fig_vr} it is apparent that there
is a significant population of physically small haloes surrounding
primary galaxies whose stellar mass corresponds to a larger halo. As
pointed out in Paper II, this is probably due to the fact that the
stellar mass varies little with halo mass for these most luminous
systems. Thus, any small scatter in stellar mass gives rise to a large
change in the value of $r_{200}$ inferred from abundance
matching. This is an important source of potential bias when trying to
measure concentrations of satellite distributions around the most
luminous primaries. In this case, a concentration derived using the
value of $r_{200}$ estimated from stellar masses will not necessarily
reflect the concentration as defined with respect to the halo mass.

Finally, as we will be investigating the colour dependence of the
results, in Fig.~\ref{fig:color} we compare the colour distributions
of galaxies in the SDSS and model galaxy catalogues. 
The distribution of SDSS galaxies in the colour-magnitude diagram is
shifted along the $g-r$ axis compared to that of the model galaxies. Thus,
while we choose $^{0.0}(g-r)_{\rm{cut}}^{\rm SDSS}=0.15-0.024M_r$ as the
line dividing the red and blue populations in the SDSS this boundary
is placed at $^{0.0}(g-r)_{\rm cut}^{\rm MS,MS-II}=-0.28-0.04M_r$ for the
model galaxy populations. These cuts are shown in the 4 panels of
Fig.~\ref{fig:color}, where the effect on the colour magnitude
distribution of including the survey flux limit can be seen for the
MS. This removes the bulk of the low luminosity galaxies that are
present in the $z=0$ snapshot and makes the resulting light-cone galaxy
population significantly more like that in the SDSS. The fraction of
the $-21<M_r<-19$ model light-cone galaxies that are defined as red is
$0.51$, the 
same as that for the SDSS. Had we adopted the SDSS colour cut, the
model would only have had a red galaxy fraction of $0.28$. For
galaxies in the magnitude range $-20<M_r<-19$, the red fractions
are 0.46 and 0.38 for the model and SDSS respectively, when using the
cuts shown in Fig.~\ref{fig:color}. Thus, apart from a
magnitude-dependent shift in the colours of galaxies, which we can
correct for with the different colour cuts, the global red galaxy
fractions are quite similar between model and SDSS data sets at
magnitudes that we will be considering for the satellite galaxies.

The MS and MS-II simulations, once populated with galaxies according to
the semi-analytic model, contain very similar galaxy
populations. Thus, it is appropriate 
to use them interchangeably depending upon which is more important:
having a large volume containing many primaries, or being able
to resolve low luminosity satellite galaxies.
However, if we want to compare results from the MS-II cube of galaxies
with those from the SDSS flux-limited survey, then 
we still need to demonstrate, using the MS, that our methods recover
from the light-cone mock surveys the same satellite distributions as
are present in the simulation cubes.

\section{Method}\label{sec:meth}

In this section we briefly review the procedure used to determine the
satellite LF and projected number density profile for SDSS, described
more fully in Papers I and II, before detailing how these
distributions are determined from the various different types of model
galaxy catalogue. The quality of the recovery of these distributions
is quantified by comparing those determined from the MS cube of
galaxies with those from the flux-limited mock light-cone surveys.

Primary galaxies are selected to have spectroscopic redshifts in the
SDSS and to have magnitudes, $M_p$, satisfying
$\satmc-\satdmbin<M_p<\satmc+\satdmbin$. We choose $\satmc=-21.0,
-22.0, -23.0$ and $\satdmbin=0.5$. Further, the primaries should be
isolated in the sense that no other galaxy within $\satdmf=0.5$
magnitudes lies within a projected distance of $2\ri$ and is
sufficiently close in redshift. `Sufficiently close' is defined as a
difference in spectroscopic redshift of less than $\satdzs=0.002$ or,
for galaxies without a spectroscopic redshift, with a photometric
redshift within $\alpha_P\sigma_P^*$, where $\alpha_P=2.5$ and
$\sigma_P^*$ is the photometric redshift error defined in Paper
II. $\ri$ represents the projected radius within which satellites may
reside, and the variable $\ro$ defines the outer edge of an annulus
within which the local background is estimated. We adopt the same
values of $(\ri,\ro)$ as in Paper~II:
$(0.3,0.6),(0.4,0.8)$ and $(0.55,0.9)~{\rm Mpc}$ for primaries in magnitude
bins $\satmc=-21.0,-22.0$ and $-23.0$ respectively.
Only galaxies brighter than $m_{\rm r}^{\rm lim}=20.5$ are considered.

Only sufficiently close galaxies in redshift are included when
counting the potential satellites within the projected radius $\ri$
and making the local background estimate from the surrounding annulus
out to $\ro$.  The background-subtracted satellite systems are stacked
for primaries in each absolute magnitude bin to provide estimates for
the mean satellite LFs and projected number density profiles of
satellites more luminous than a particular absolute magnitude, as
described more fully in Papers~I \& II.

The procedure described above is applied to the SDSS itself and also
to the MS redshift space light-cone mock catalogue. However, different
estimation procedures are used for the light-cone mock with real space
(rather than redshift space) galaxy coordinates and the galaxy
populations in the simulation cubes.  With the real space positions,
it is possible to define isolated primaries as having no bright
neighbours within a sphere of radius $2\ri$. The satellites within a
sphere of radius $\ri$ can also be determined without any need for
background subtraction.  Using only the real space satellites, the
estimation of the satellite luminosity function is straightforward.
For the $j$th magnitude bin, the average satellite LF is estimated
using

\begin{equation}
    \overline{N}^{\rm real\ sat}_{j}= \frac{\Sigma_iN^{\rm real\ sat}_i(M_j)}
    {N_j^{\rm real\ prim}},
\end{equation}
where $N^{\rm real\ sat}_i$ is the number of satellites around primary
$i$ and $N_j^{\rm real\ prim}$ is the number of primaries contributing
to the $j$th bin of the LF. 

While we need no correction for interlopers in real space, the
process of estimating the mean projected number density of satellite galaxies
is such that a fair comparison with light-cone data still requires
the subtraction of a background estimated from an outer area.
Hence, the projected number density profiles of satellites brighter
than $M^{\rm trun}$ in real space are determined using
all galaxies within a cylinder of projected radius $\ro$ and length
$2\ro$, centred on the primary. These galaxies are projected onto a 
plane and provide the sources for the potential satellites/background
for projected radii less/greater than $\ri$. The 
projected number density profile is determined using
\begin{equation}
  \overline{\Sigma}_i(r^{\rm ann}_j)
     =\frac{\sum_i N_{ij}(M^{\rm trun})}{\sum_i A_{ij}^{\rm p}} 
- \frac{\sum_i N^{\rm bck}_i}{\sum_i A^{\rm outer}_i  },
\end{equation}
where $N_{ij}$ is the number of galaxies brighter than
$M^{\rm trun}$ within a projected distance $\ri$ of the $i$th primary
and in the $j$th projected annulus,
and $N^{\rm bck}_i$ is the corresponding number of galaxies in the
projected outer
annulus, $\ri < r < \ro$. $A_{ij}^{\rm p}$ is the 
area contributed by the $i$th primary to the $j$th annulus for the detection
of satellites brighter than $M^{\rm trun}$,
\begin{equation}
 A_{ij}^{\rm p}(M^{\rm trun})=\left\{
\begin{array}{ll}
  A_{ij} &  M^{\rm trun} < M_i^{\rm{lim}} \\
  0 &   M^{\rm trun} > M_i^{\rm{lim}} \ ,
\end{array}\right.
\end{equation}
where $A_{ij}$ is the area of the $j$th annulus surrounding the $i$th primary
and $M_i^{\rm lim}$ is the absolute magnitude that corresponds to the apparent
magnitude limit of the mock catalogue. $A_i^{\rm outer}$ is the
corresponding area in the outer annulus surrounding the $i$th primary.

\begin{figure}
\centering
\includegraphics[width=78mm]{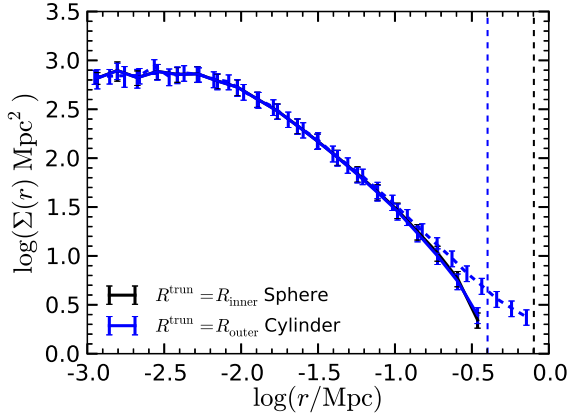}
\caption{The comparison of real space projected number density
    profiles estimated by different methods. The black line is the
    profile estimated by projecting the satellites within spheres of
    radius $\ri$ around 
    primaries. The dashed blue curve shows the projected number density
    profile of all galaxies within a line-of-sight distance $\ro$ of a
    primary. The solid blue curve is the background-subtracted case,
    where the background is estimated from the outer annulus with
    $\ri<r<\ro$. Vertical dashed lines show $\ri$ and $\ro$.}
\label{fig:bg}
\end{figure}

The comparison of the projected number density profiles, before and after
background subtraction, with that formed by simply projecting the
galaxies within a sphere of radius $\ri$ of the primary galaxy is
shown in Fig.~\ref{fig:bg}. The 
results indicate that the projected profile after subtracting the background
very accurately recovers that
estimated directly from the galaxies within the inner area
($r<\ri$). The impact of the background subtraction is small and
limited to radii near to $\ri$.
This establishes that the method for calculating the background
subtracted projected number density profile from a real space light-cone
survey provides an unbiased estimate of that produced when only satellites
within a 3D distance $\ri$ are used. We can now compare these real
space profiles with those from redshift space light-cones.

\begin{figure}
\includegraphics[width=80mm]{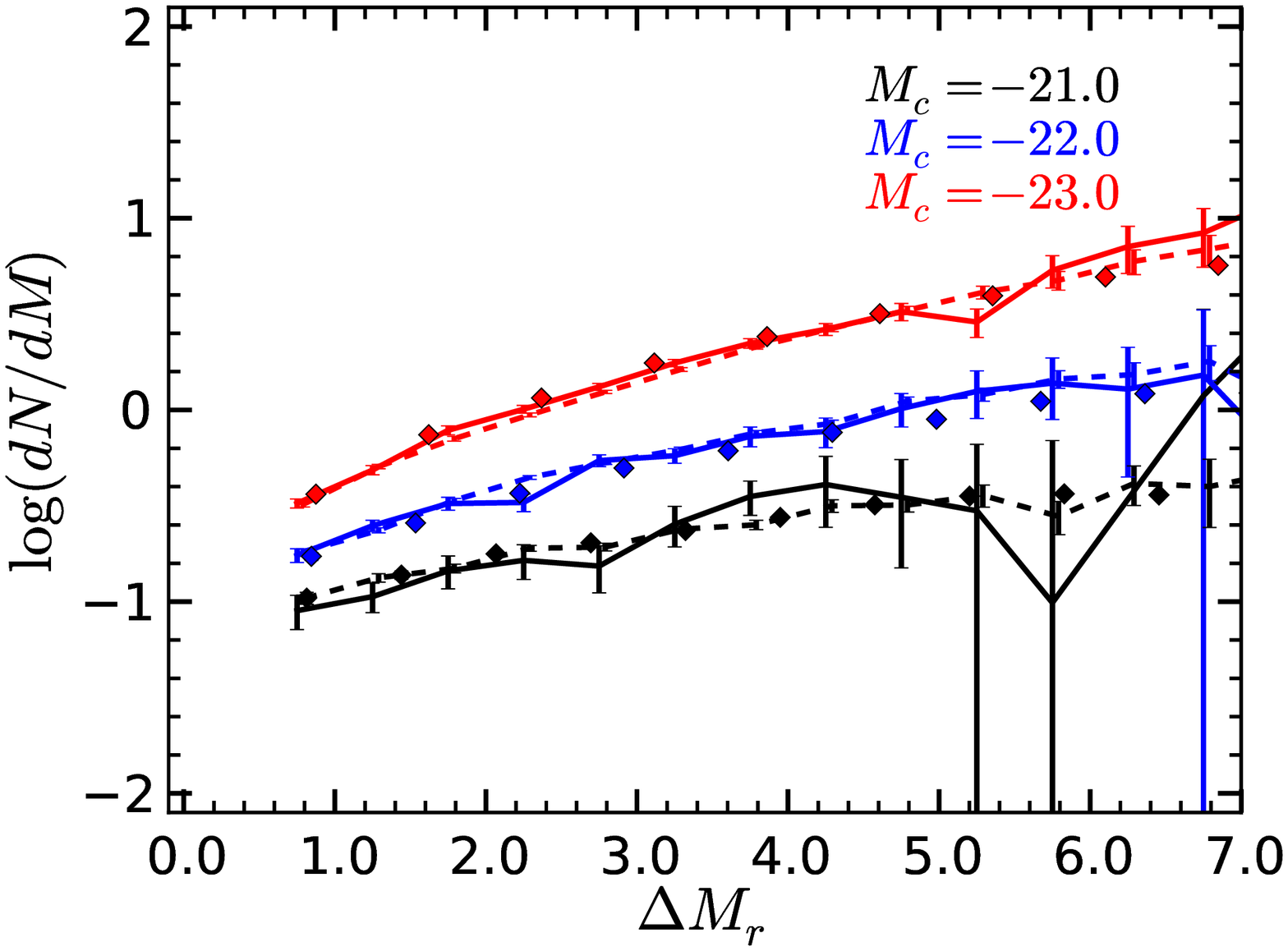}
\includegraphics[width=80mm]{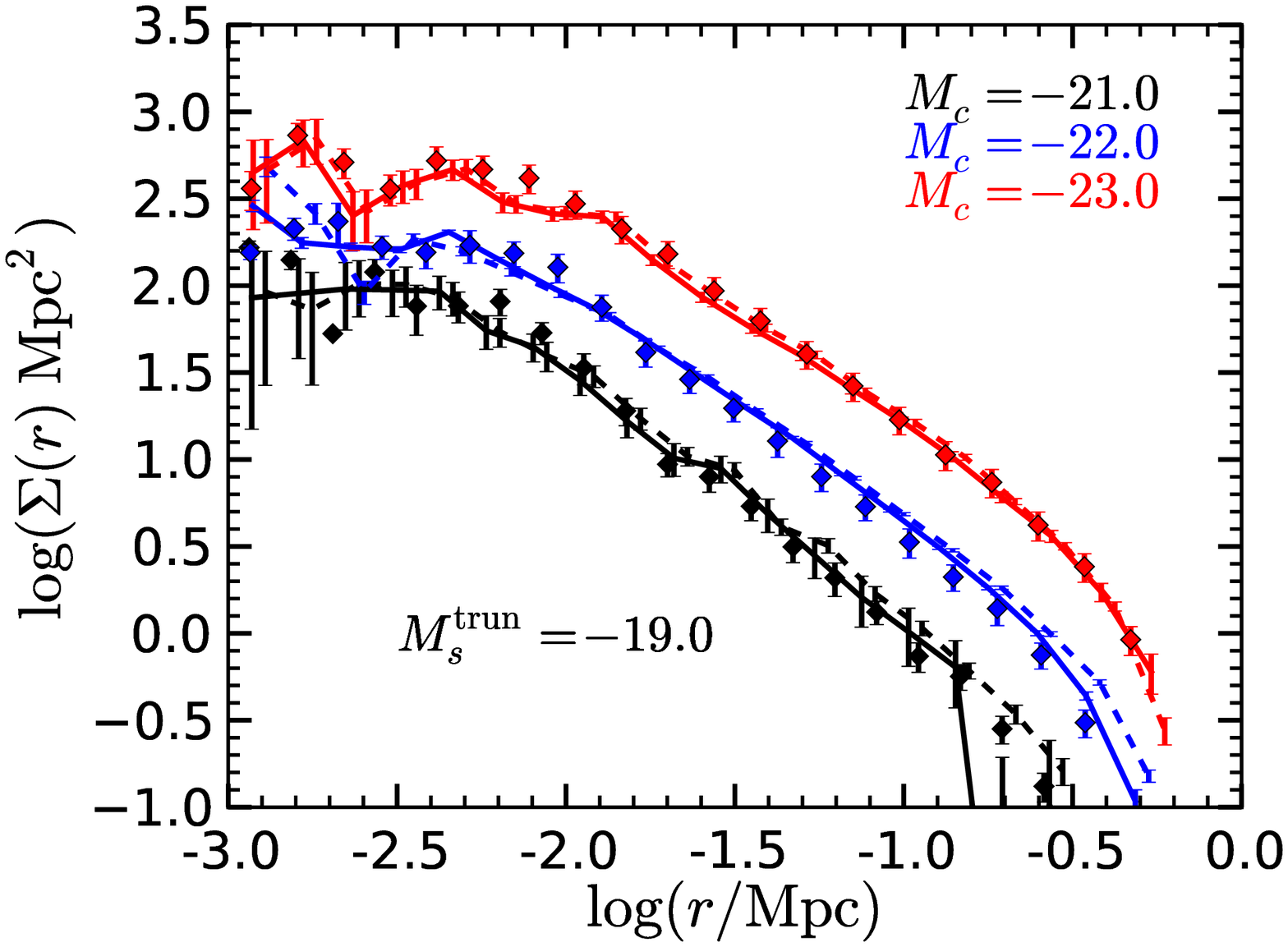}
\caption{Satellite LFs (top) and number density
    profiles (bottom) estimated from light-cone mock catalogues and
    directly from the MS cube. The results for primary
    magnitude bins centred at 
    $\satmc=-21.0, -22.0, -23.0$ are shown in black, blue and red
    respectively. Solid and dashed lines correspond to results from
    redshift and real space light-cones respectively, whereas the
    points show the results for the whole volume in the simulation cube.}
\label{fig:fig2}
\end{figure}

Having described the three different types of model galaxy catalogues
made using the MS and how the
satellite distributions are determined from each of them, we are now
in a position to test the accuracy of our estimation
procedure. Fig.~\ref{fig:fig2} shows the satellite LF and projected
number density profiles estimated from each of the model galaxy
catalogues: simulation cube, real space light-cone and redshift space
light-cone. The agreement between the two different satellite
distributions measured from all three catalogues at $\satmc=-21.0,
-22.0$ and $-23.0$ provides strong
support that our results are unbiased, and that our technique for
background subtraction is appropriate. Given that the isolated
primaries are a specially selected subset of the differentially
clustered galaxy population, one would expect that a local, rather
than global, background subtraction would be appropriate.
The fact that the results from the MS light-cone mock match well with
those from the MS cube of galaxies suggests that we can use
results from the MS-II simulation cube at $z=0$ when comparing with low
luminosity galaxies in the SDSS DR8.

\begin{figure}
\centering
\includegraphics[width=78mm]{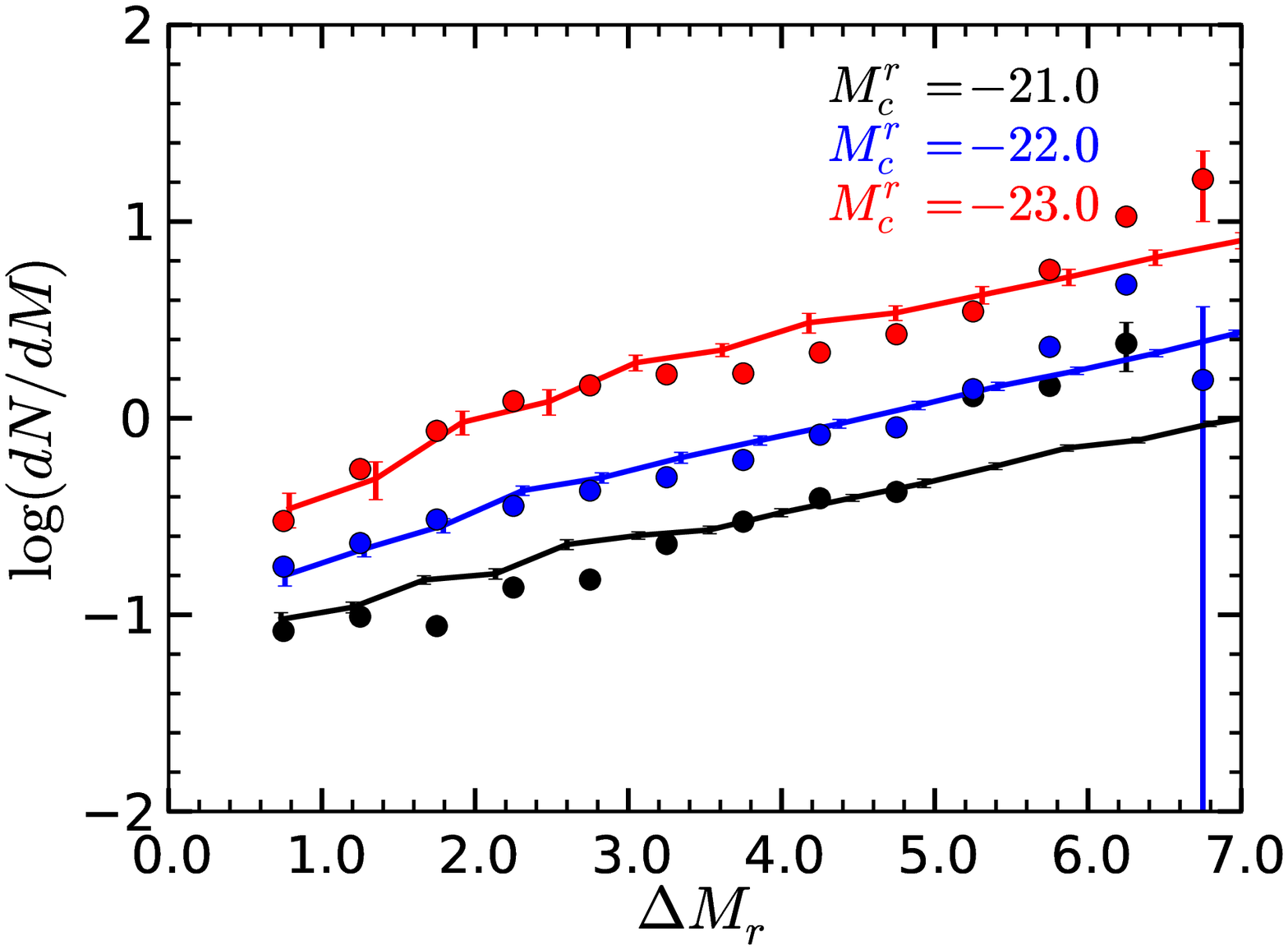}
\includegraphics[width=78mm]{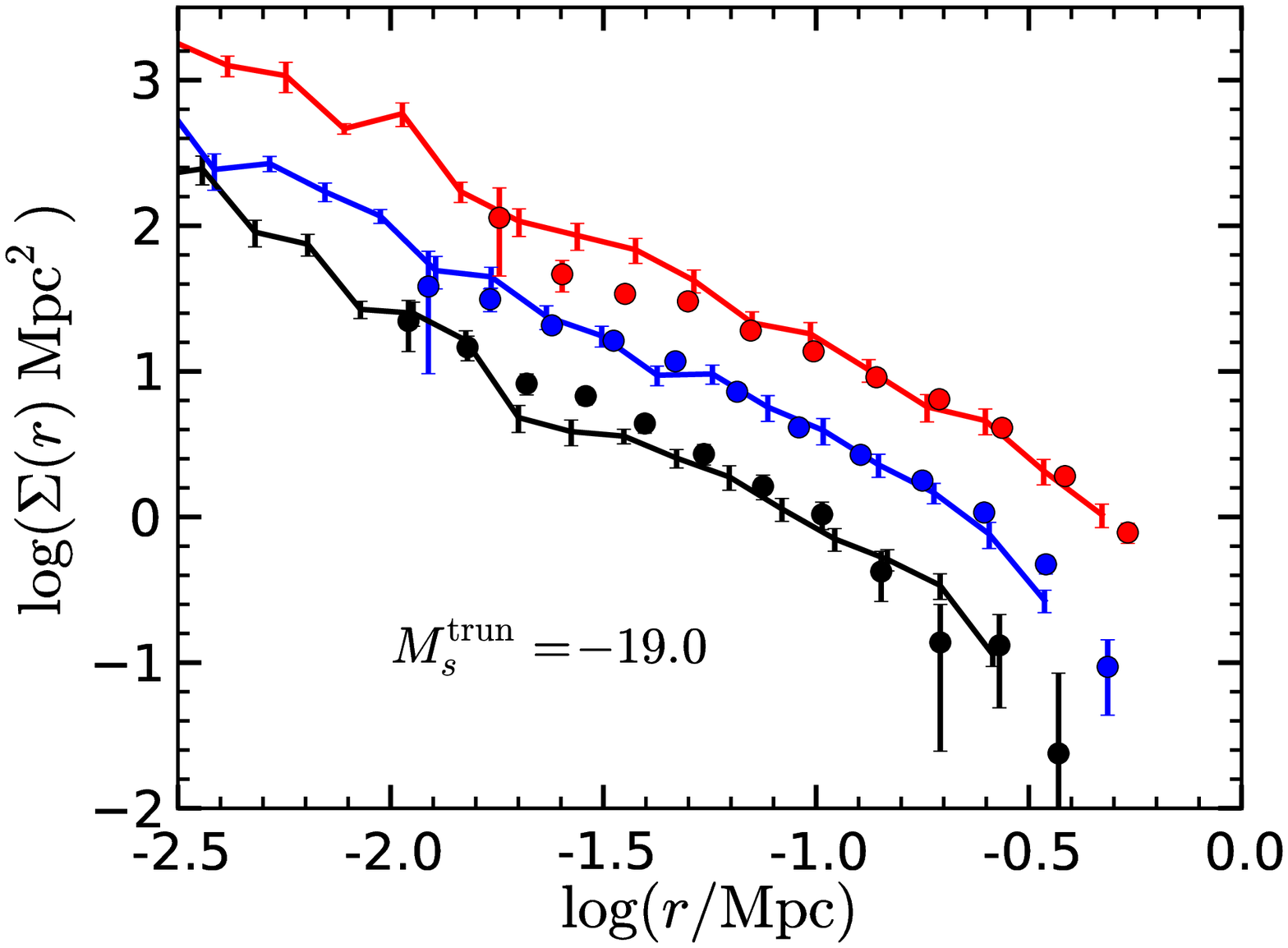}
\includegraphics[width=78mm]{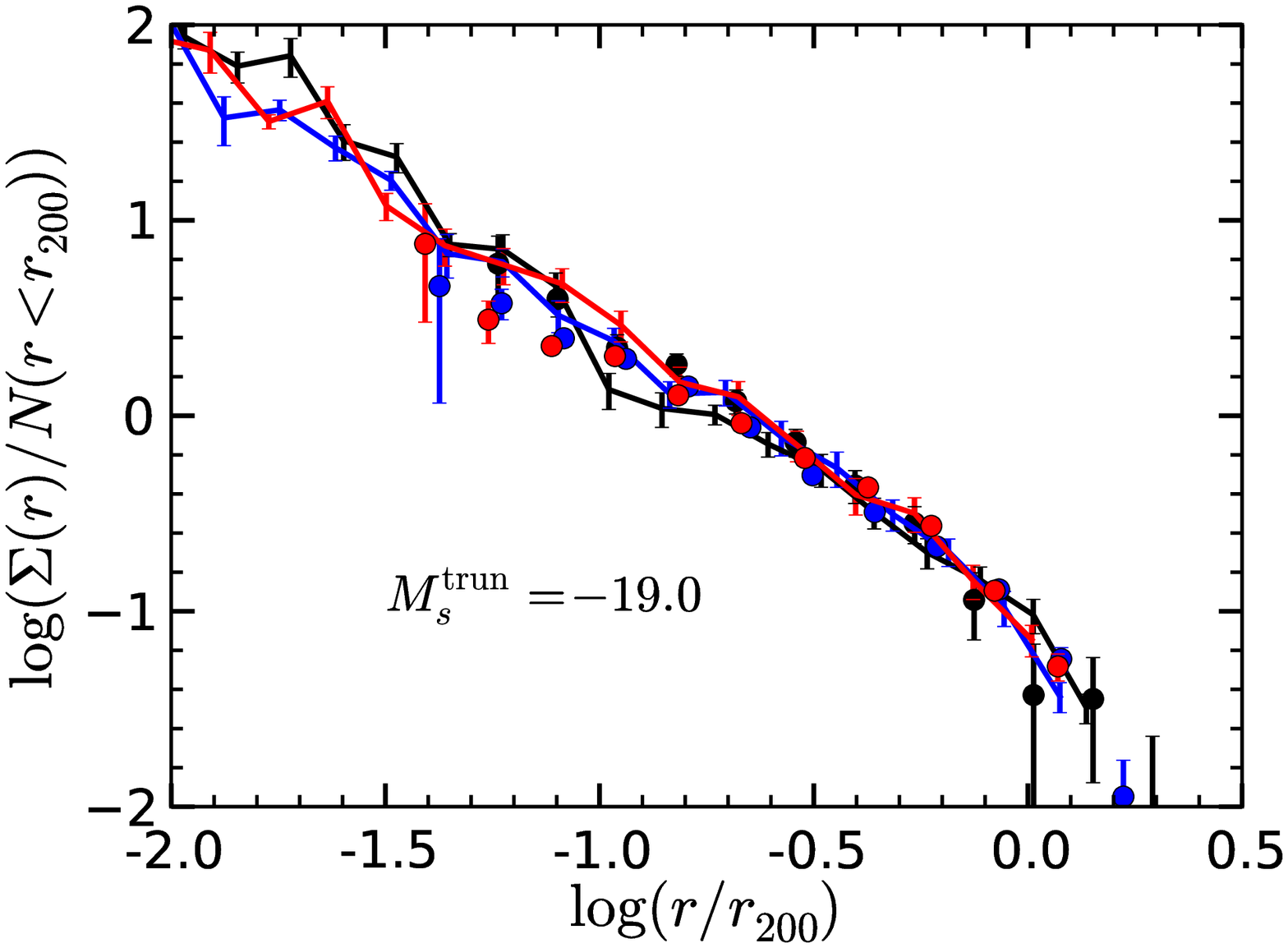}
\caption{Model (MS-II, solid lines) and SDSS (points) satellite LFs (top),
    projected number density profiles (middle) and normalized profiles
    (bottom). The results for primary luminosity bins $\satmc=-21.0, -22.0,
    -23.0$ are shown in black, blue and red, respectively.}
\label{fig:general}
\end{figure}

\begin{figure}
\includegraphics[width=78mm]{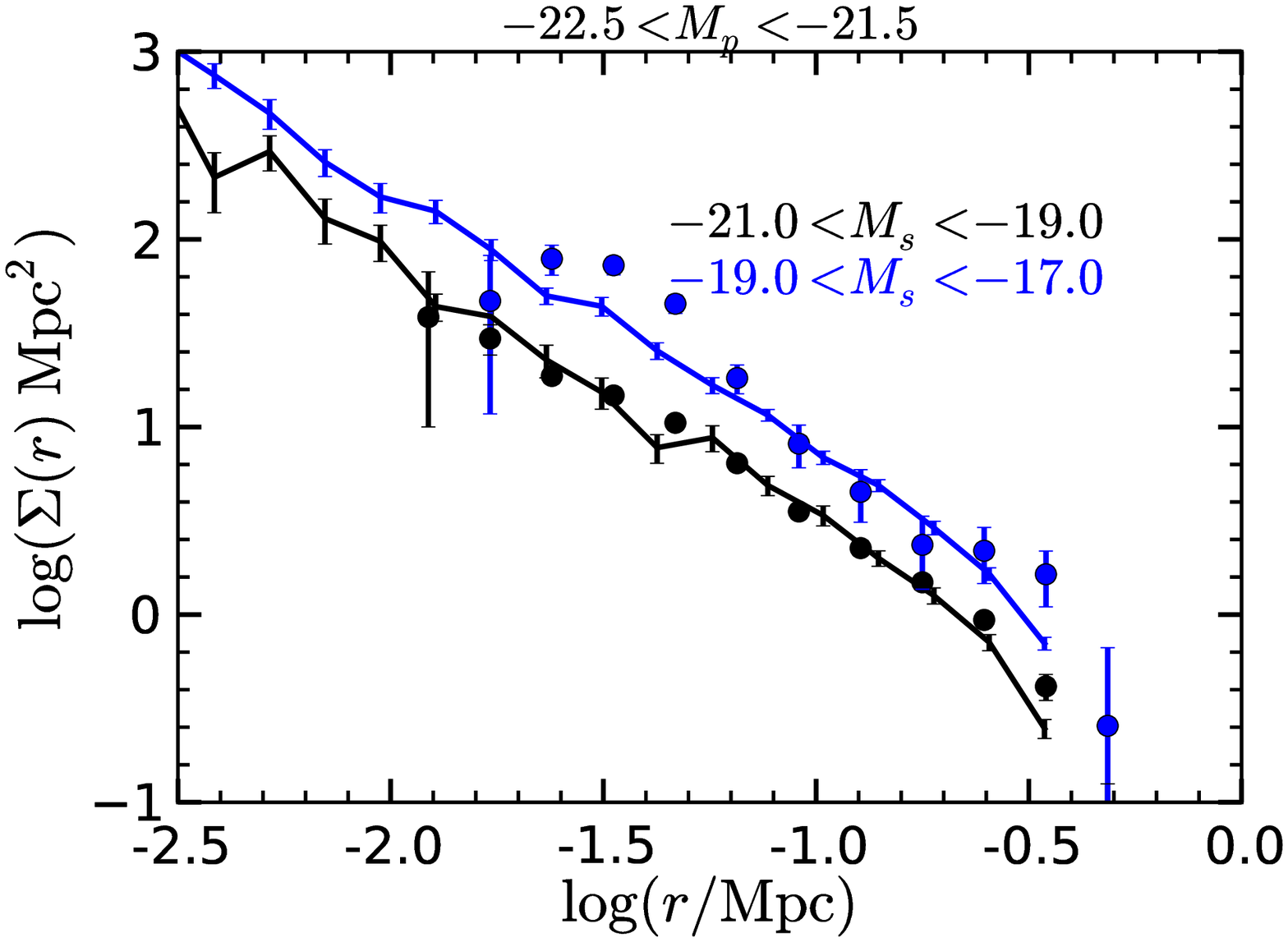}
\includegraphics[width=78mm]{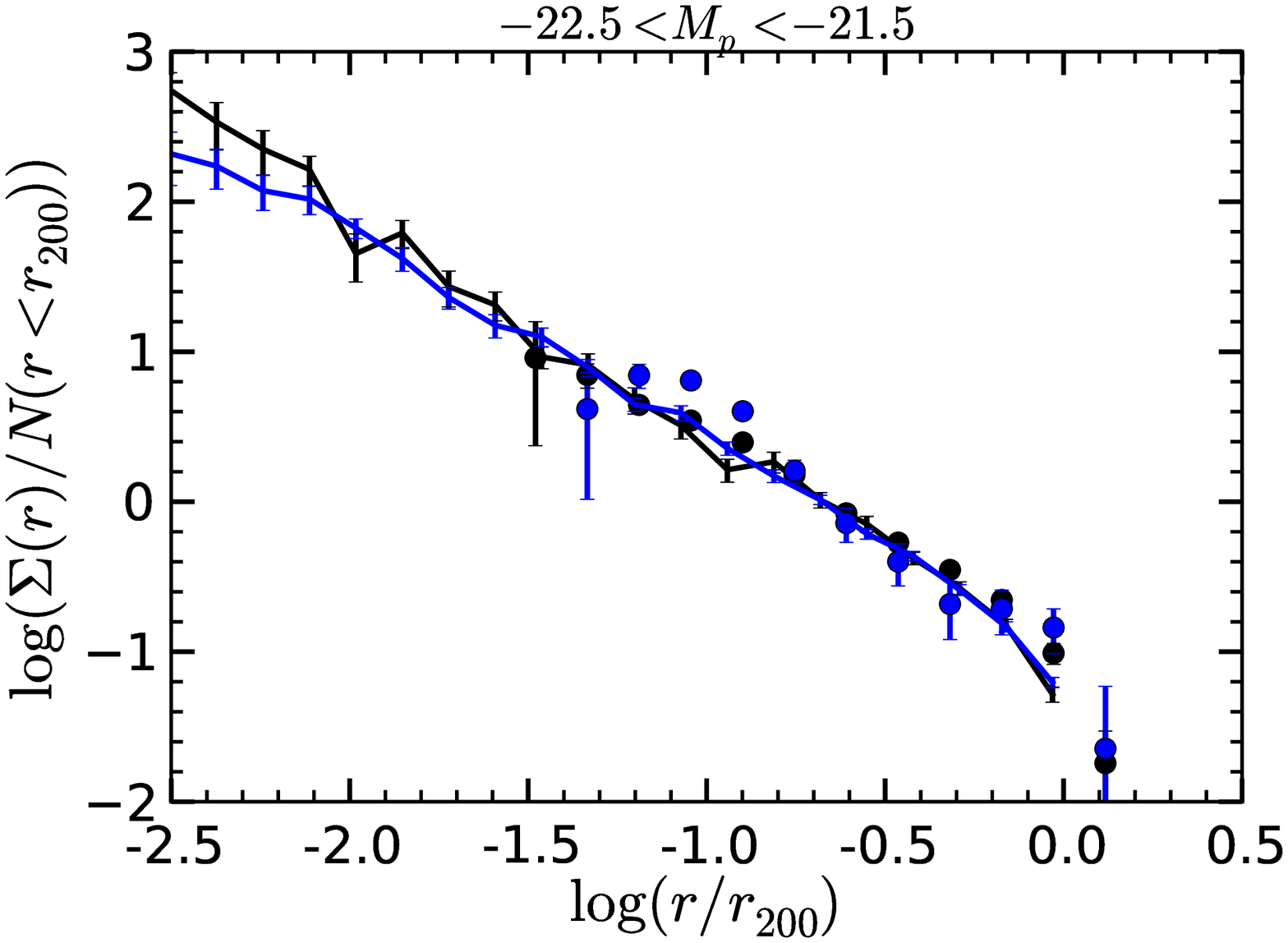}
\includegraphics[width=78mm]{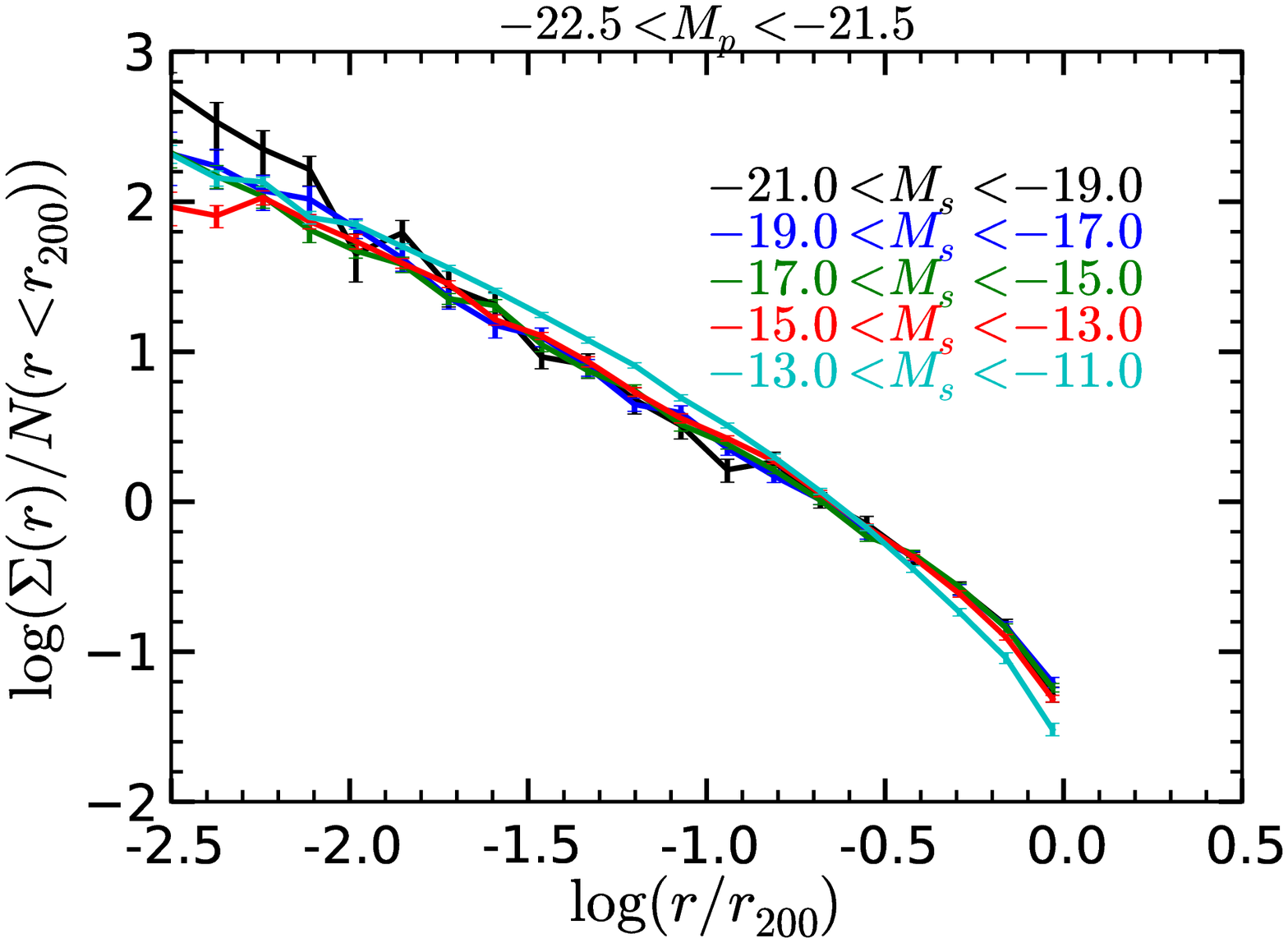}
\caption{The dependence of the scaled and unscaled satellite density profiles on
    satellite luminosity for primaries in the range $-22.5 <
    M_{\rm p} < -21.5$. Unscaled (top) and scaled (middle) profiles of
    satellites are shown for two different luminosity bands:
    $-21.0<M_s<-19.0$ (black) and $-19.0 <M_s<-17.0$ (blue). The solid
    lines and points correspond to results from the model and SDSS respectively.
    The bottom panel shows scaled profiles for lower luminosity
    satellites in the model.}
\label{fig:pro}
\end{figure}

\section{Results}

Having verified that the model galaxies have similar
distributions of luminosity and colour to those in the SDSS and that our
local background subtraction procedure produces unbiased estimates of
the satellite LF and projected number density profile, we now
compare the model and observed satellite systems in more detail.

\subsection{Dependence on primary and satellite luminosity}
The top panel of Fig.~\ref{fig:general} shows the satellite LFs for the primary
magnitude bins $\satmc=-21.0, -22.0, -23.0$ estimated from both the SDSS and
MS-II mock data. For $\Delta M_r<5$, the model and SDSS satellite
luminosity functions generally agree well. However, there is a steepening
of the LF for lower luminosity satellite galaxies in the SDSS that is
not present in the model. The MS-II was used for this comparison, so a lack
of numerical resolution should not be responsible for this deficit.

The middle panel of Fig.~\ref{fig:general} shows the projected number
density profiles of satellites brighter than $\msatt=-19.0$ for the
different primary samples. This limit is chosen, following Paper II,
to ensure that the profiles for different luminosity bins are measured
from a large enough sample.  In all cases the profile approximately
follows $\Sigma(r)\propto r^{-1.5}$, with the amplitude reflecting the
fact that more luminous primaries host more satellites. For most
radii, the model reproduces the amplitude observed in SDSS. In detail
though, there is a factor 2 difference at $\sim 30~{\rm kpc}$ for both
$\satmc=-21.0$ and $-23.0$. The model underproduces satellites at this
radius for the least luminous primaries and overproduces them for the
most luminous primaries.

Once rescaled in both radius and projected number density, as shown in
the bottom panel of Fig.~\ref{fig:general}, the model and SDSS
profiles line up well outside the region at $0.05\lsim r/r_{200} \lsim
0.1$. As noted in Paper II, the slight deficit of satellite galaxies
in the inner regions of the SDSS for $\satmc=-23.0$ primaries,
relative to the model, may reflect difficulties identifying low
luminosity satellites in regions where the background light
subtraction is significant \citep{aih11}. This had the effect of
slightly decreasing the fitted concentration of satellites around the
most luminous primaries relative to the low luminosity primary bins.

The dependence of the satellite projected number density profile on
satellite luminosity for primaries with $-22.5<M_p<-21.5$ is shown in
Fig.~\ref{fig:pro}. Satellites are split into two bands of luminosity:
$-21.0 < M_s< -19.0$, representing the objects that contributed to the
profile in Fig.~\ref{fig:general}, and $-19.0 < M_s<-17.0$, showing
the behaviour of lower luminosity satellites.  The top and middle
panels show the satellite projected number density profiles before and
after scaling respectively. While generally in good agreement, there
is an excess of lower luminosity satellites in the SDSS relative to
the model at $\sim 30~{\rm kpc}$.

The `bump' in the projected number density profile in the SDSS data is
present only for the lower luminosity satellites. One is therefore
tempted to ask if it is present at any satellite luminosity in the
mock catalogues. This question is answered in the bottom panel of
Fig.~\ref{fig:pro}, where no comparable deviations are seen for
satellites with $M_s<-13$, which have indistinguishable profiles from
those of brighter satellites. The profile for satellites with
$-13<M_s<-11$ shows a slight change in shape over a large range of
scales relative to the other sets of satellites, but nothing as
pronounced as is seen in the SDSS.

\subsection{Dependence on primary and satellite colour}

\begin{figure}
\includegraphics[width=78mm]{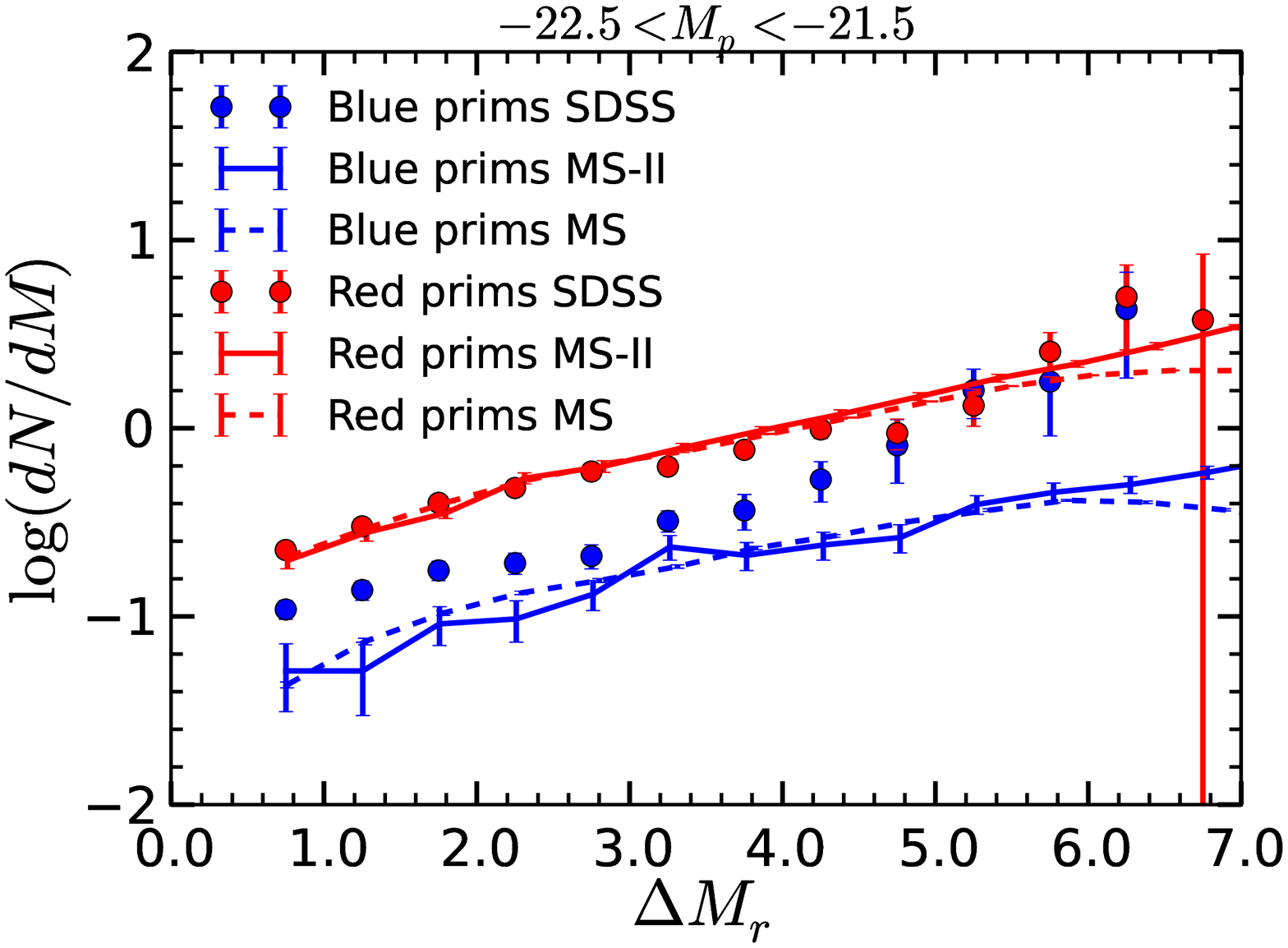}
\includegraphics[width=78mm]{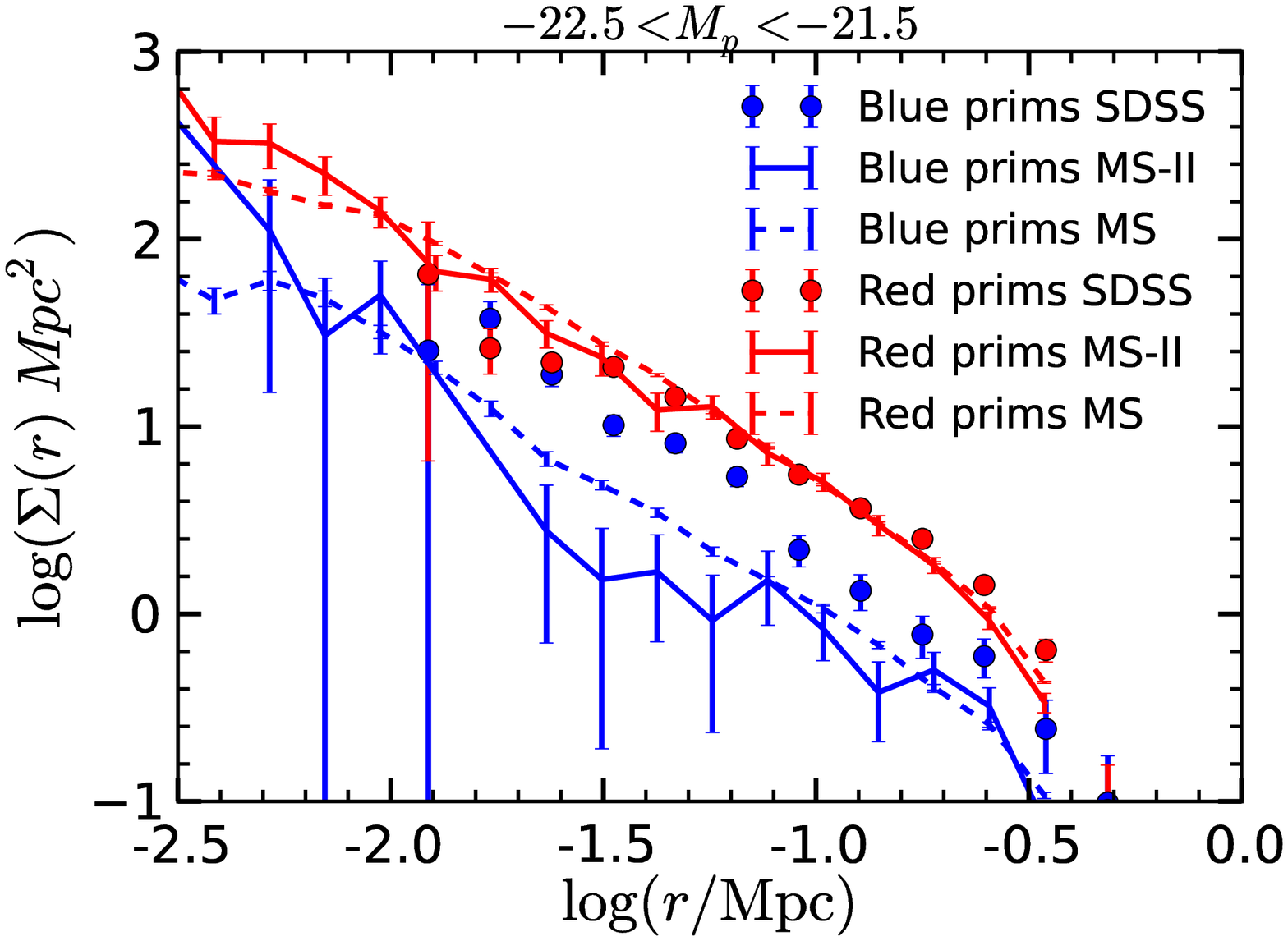}
\includegraphics[width=78mm]{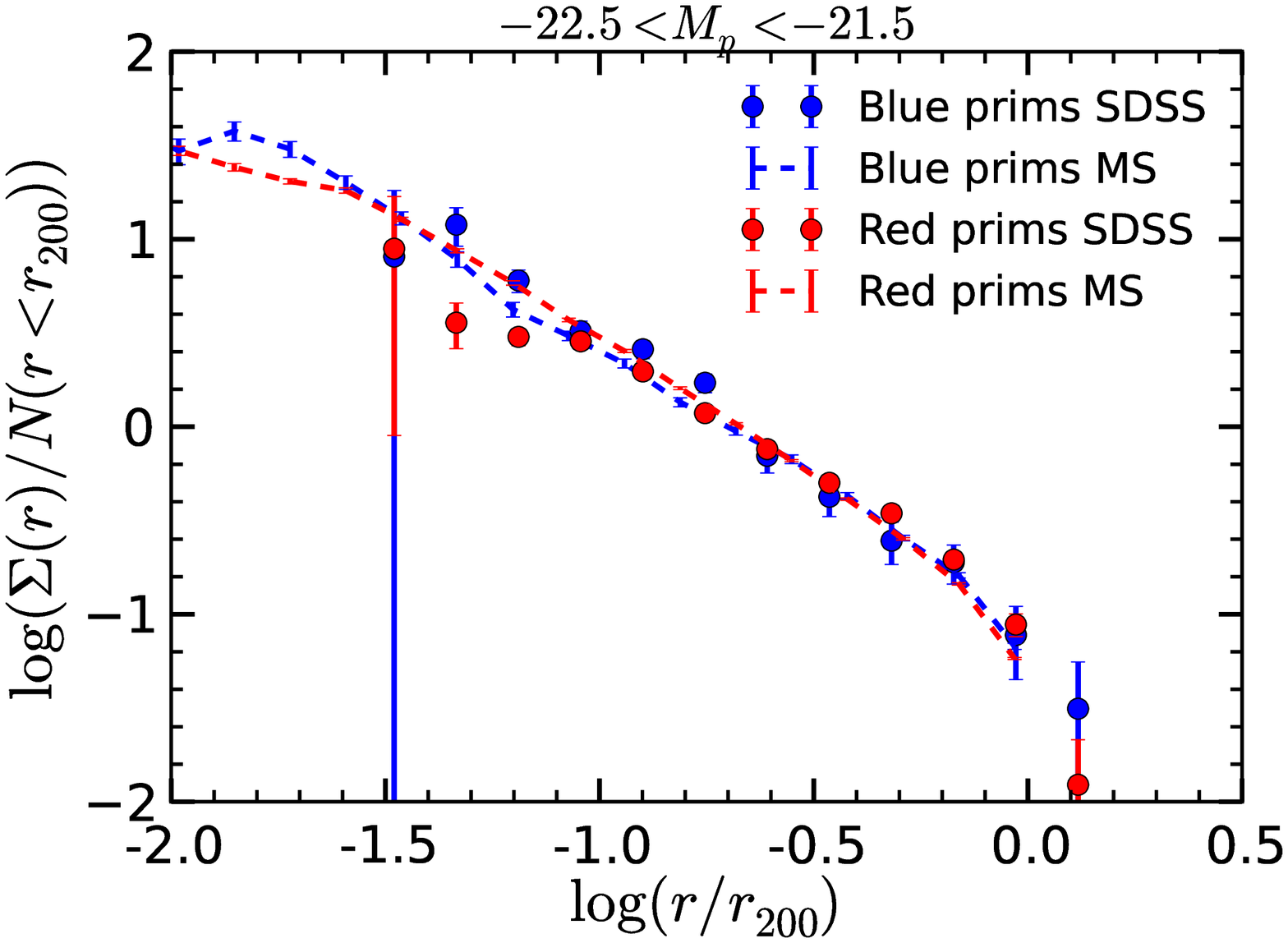}
\caption{The dependence on primary galaxy colour of the satellite LF
    (top), unscaled (middle) and scaled (bottom) number density profiles
    for primary galaxies of magnitude $-22.5<M_p<-21.5$. 
    Solid and dashed lines show model results for
    primaries in the MS and MS-II respectively, whereas the points are
    for SDSS. All profiles are for
    satellites more luminous than $M_s=-19.0$.}
\label{fig:prim_color}
\end{figure}

\begin{figure}
\includegraphics[width=78mm]{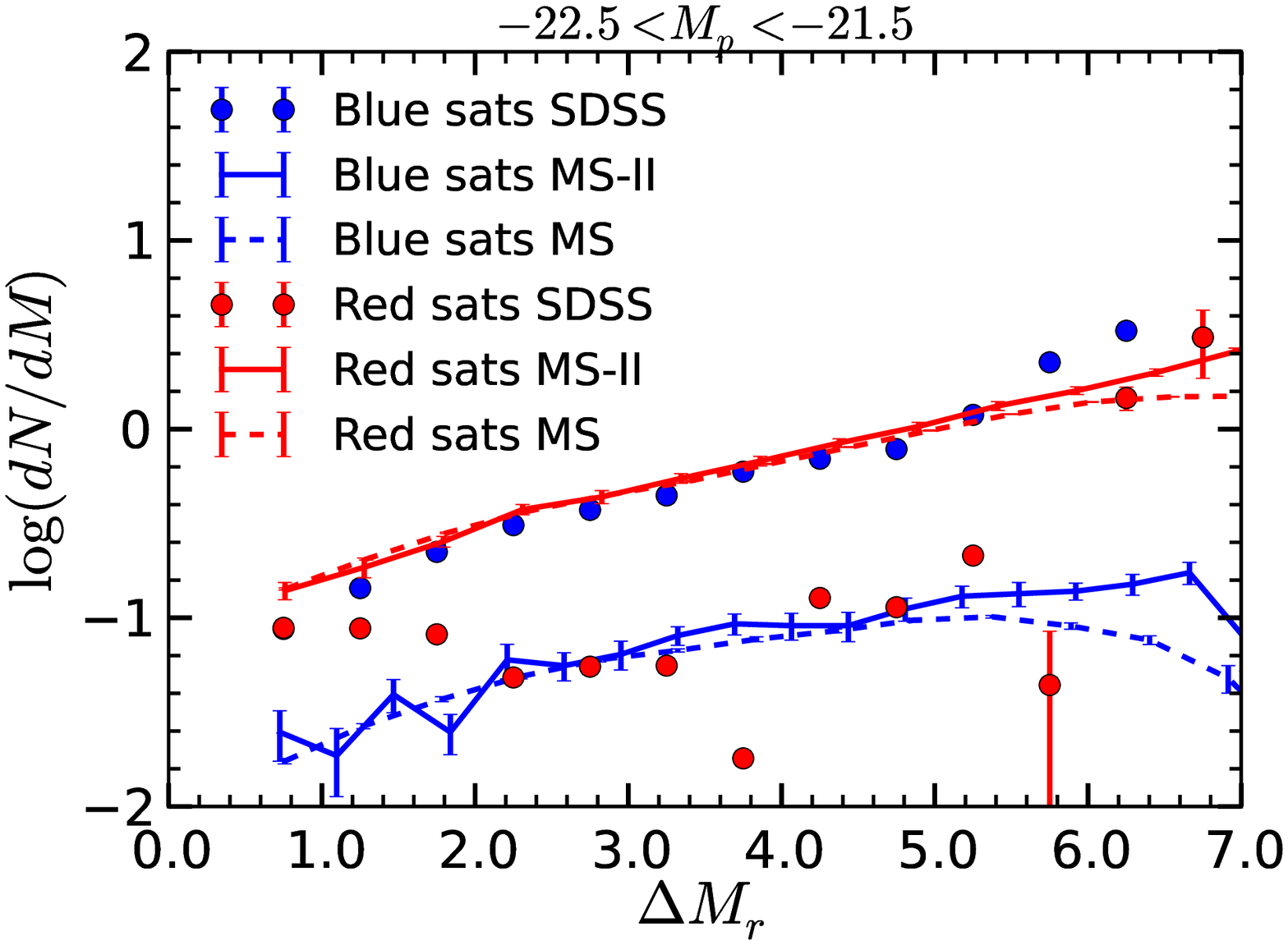}
\includegraphics[width=78mm]{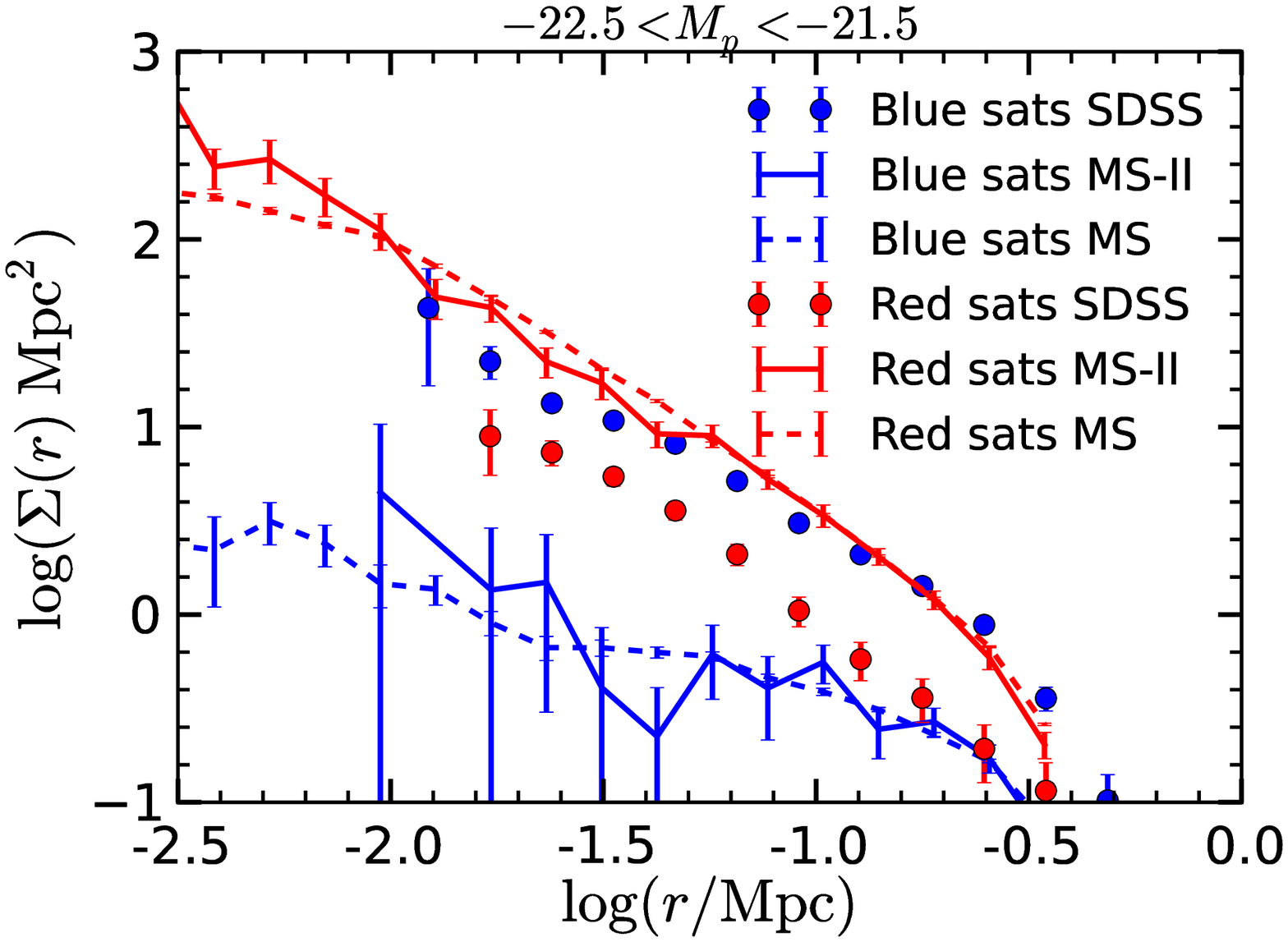}
\includegraphics[width=78mm]{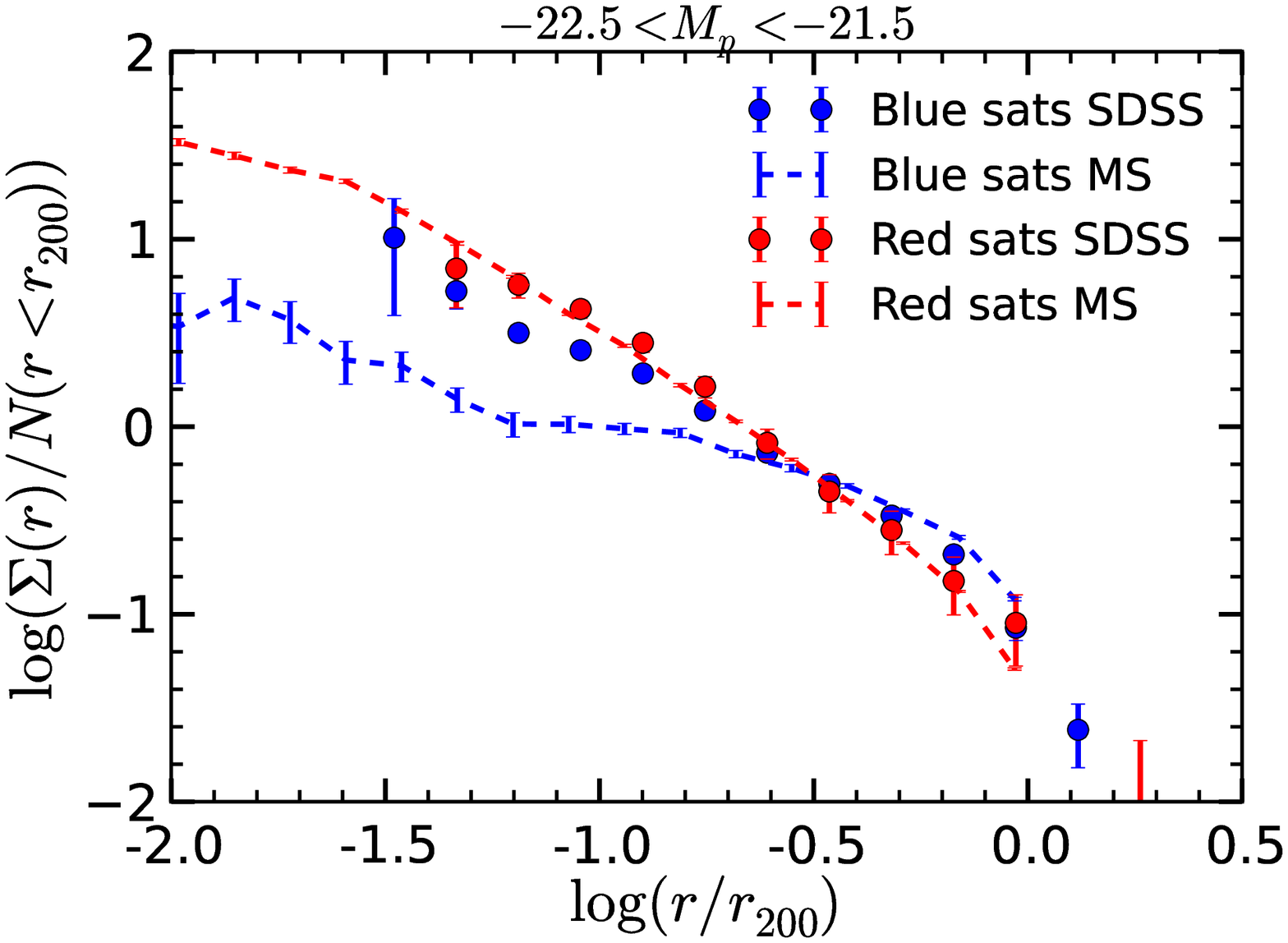}
\caption{Dependence of satellite distributions on the colour of the
    satellite galaxy. Symbols and line types have the same meaning as
    in Fig.~\ref{fig:prim_color}.}
\label{fig:sat_color}
\end{figure}

As a further test of the galaxy formation model, which has so far largely
succeeded in reproducing the satellite LF and projected number density
distributions, we now split the primaries and satellites by
colour using the slightly different colour cuts described in
Section~\ref{ssec:model} for the model and SDSS.

Fig.~\ref{fig:prim_color} shows how the satellite luminosity function
and projected number density profile depend upon the colour of the
primary around which they are being measured, and compares the results
from the model and the SDSS for primaries with $-22.5<M_p<-21.5$. For
the adopted colour cuts, the vast majority of primaries are classified
as red in both the model and SDSS. Given that when not split by colour
the model produces a good match to these satellite distributions, it
is no surprise that the satellite populations around red primaries
agree well between model and observations. However, the blue primaries
in the mock catalogues are deficient in satellites by a factor of
$2-3$. The excess satellites around SDSS primaries span the range of
luminosities and radii being considered here, with a slight tendency
to be at lower luminosity and nearer to the primary than for the
satellites present around the model primaries. The scaled profiles in
the bottom panel of Fig.~\ref{fig:prim_color} show that satellites
around blue SDSS primaries are slightly more concentrated than those
around model primaries. 

In the top two panels of Fig.~\ref{fig:prim_color} results for both
the MS and MS-II are shown. Because of its limited resolution the LF
in the MS becomes incomplete at $\Delta M\sim 5.5$, but the LF in the
MS-II is well resolved down to $\Delta M\sim 7$. On the other hand, in
the smaller volume of the MS-II, there is a relatively small number of
primaries and, as a result, the projected number density profile of
satellites brighter than $M_s=-19$ is noisy (and it is therefore
omitted in the lower panel of the figure).  In the regions where both
simulations are well resolved and sampled, their results are
consistent.

The satellites themselves can be divided into red and blue subsets and
their distributions around primaries with $-22.5<M_p<-21.5$ are shown
in Fig.~\ref{fig:sat_color}. Once again both MS and MS-II results are
shown in the top two panels, while the noisy MS-II results are not
presented for the scaled projected number density profiles in the bottom panel.
The fact that the colour cuts in the model and SDSS samples yield
completely different red and blue fractions is immediately apparent in
the satellite LFs, with the model satellite systems dominated by red
satellites to a coincidentally similar extent as the blue satellites dominate
around SDSS primaries. For the most luminous satellites, the SDSS blue
and red fractions converge, whereas this does not happen for the
model.

The shape of the projected density profiles of red satellites are very
similar for the model and SDSS systems, once the difference in
amplitude has been scaled away, as shown in the bottom panel of
Fig.~\ref{fig:sat_color}. However, this is not the case for the blue
satellites, which have a much less concentrated distribution around
model primaries. For real systems of blue satellites, the distribution
is only slightly less concentrated than it is for the red satellites.

\section{Discussion}

The comparison of galactic satellite systems in the model with those
in the SDSS shows that the dependence of the satellite distributions
on primary and satellite luminosity is captured quite well by the \gf
model. This is a non-trivial success of the model since its parameters
were adjusted merely to match the global K-band LF of galaxies, with
no direct reference to satellite systems.  There is an excess of very
low luminosity satellites around SDSS primaries relative to the model,
and the projected number density profiles are up to a factor 2
discrepant within $\sim 30~{\rm kpc}$, but the agreement is generally
good.

\cite{tal12} also studied the radial distribution of satellite systems
around bright primary galaxies using SDSS data. Their primaries were
LRGs at $0.28<z<0.40$ with no isolation criteria applied and hence
often in groups of bright galaxies, and thus are different to those
studied here in a few respects that may well be important. They found
the projected number density was well fitted by a combination of a
projected NFW profile \citep{1996ApJ...462..563N,NFW_97} for large
radii and a central steeper profile that follows the stacked light
profile of the LRGs. This central bump is similar to that seen for the
lower luminosity satellites around primary galaxies shown in
Fig.~\ref{fig:pro}, but is, in contrast, most apparent in the high
luminosity satellites around the LRGs.

The differences between the model and SDSS satellite systems are greater when
the colour of the satellites is considered. Even the distribution of galaxies in
the colour-magnitude plane shows that the model has too high a fraction of low
luminosity red galaxies relative to the SDSS. The model blue satellites are both
significantly depleted and very much less concentrated relative to either blue
or red SDSS satellites, which have a projected number density profile like that
of the red model satellites. These pieces of evidence point to the model being
too ready to convert low luminosity blue galaxies to red ones. 

\begin{figure}
\includegraphics[width=77mm]{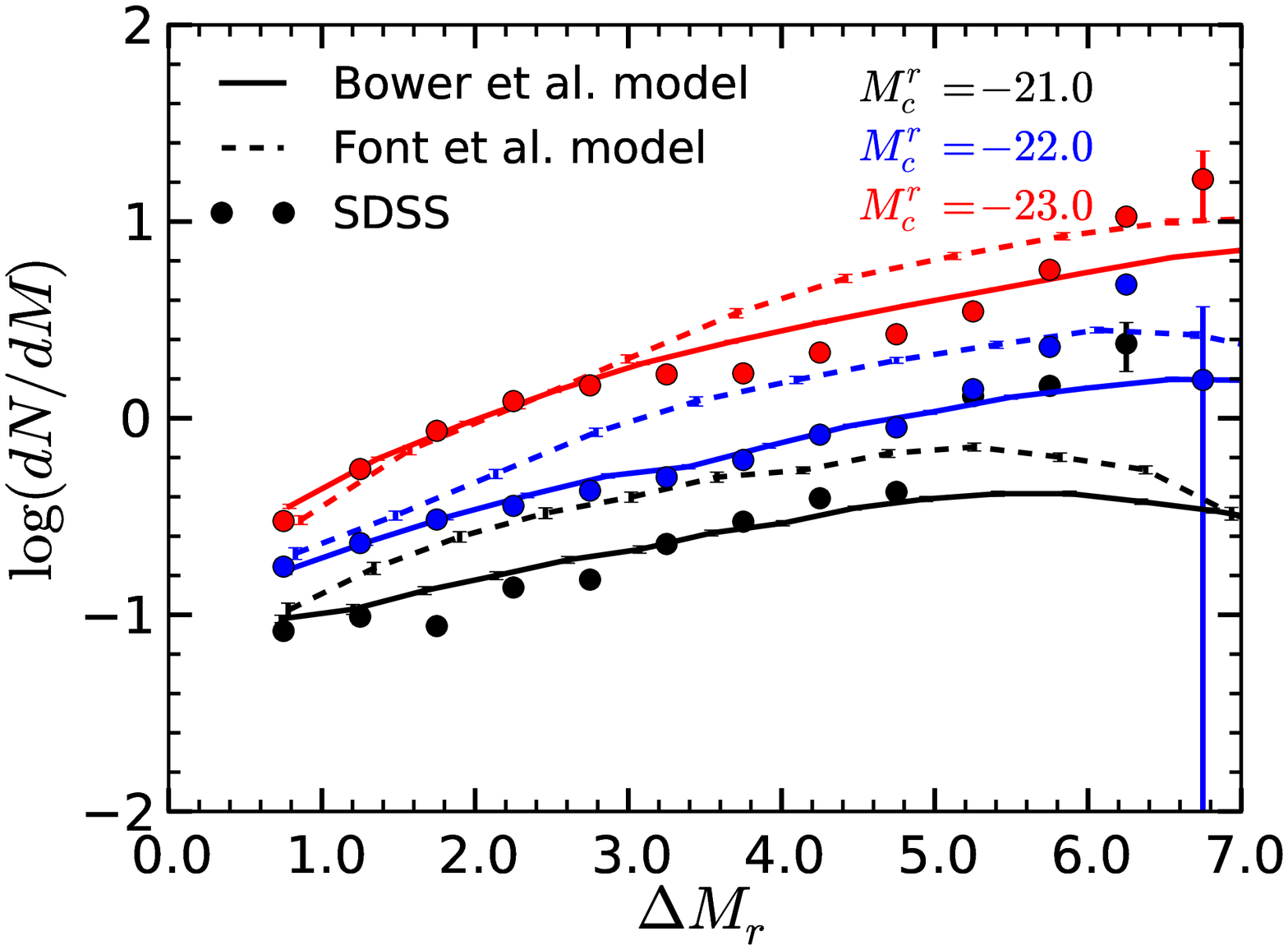}
\includegraphics[width=77mm]{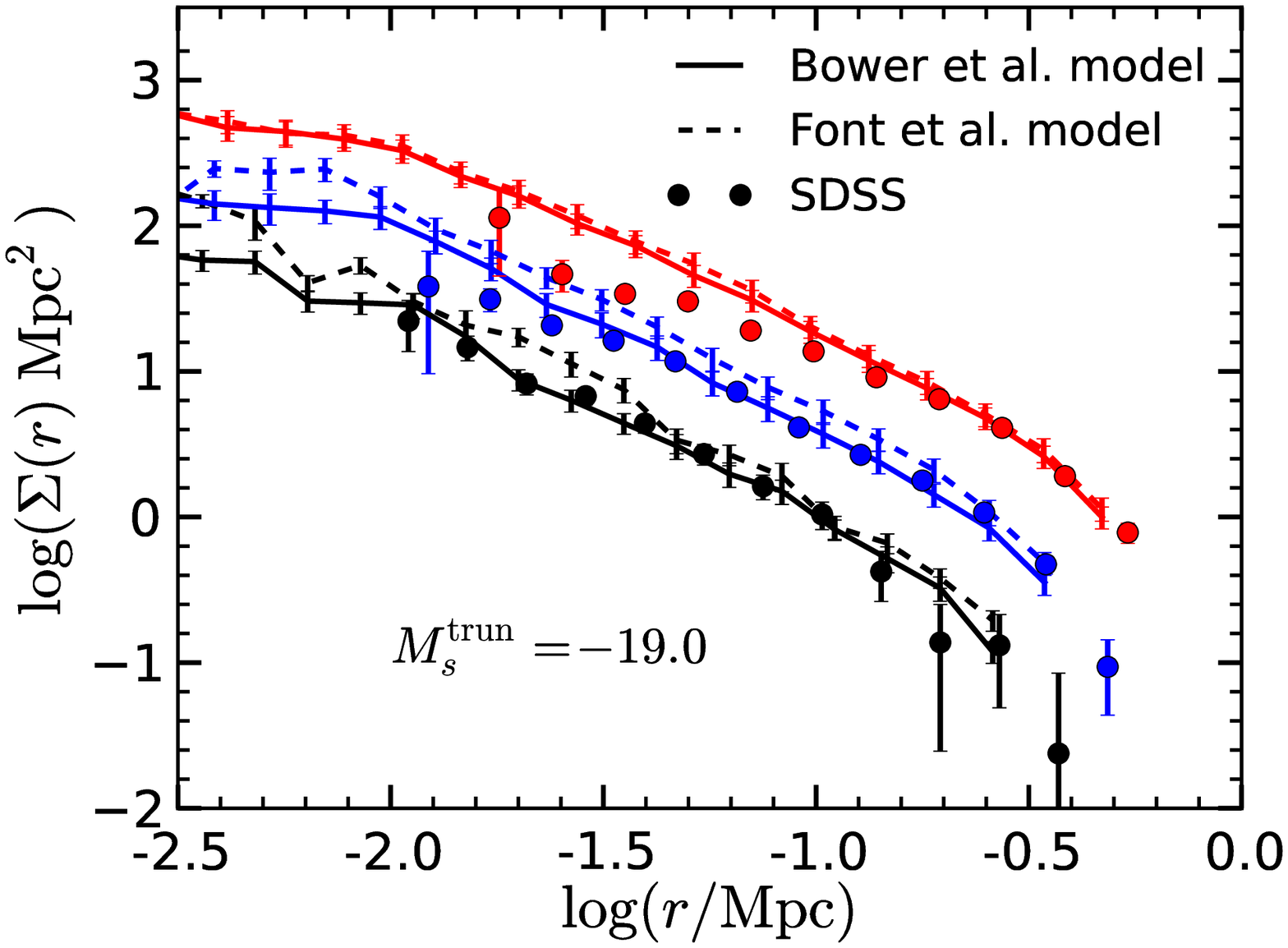}
\includegraphics[width=77mm]{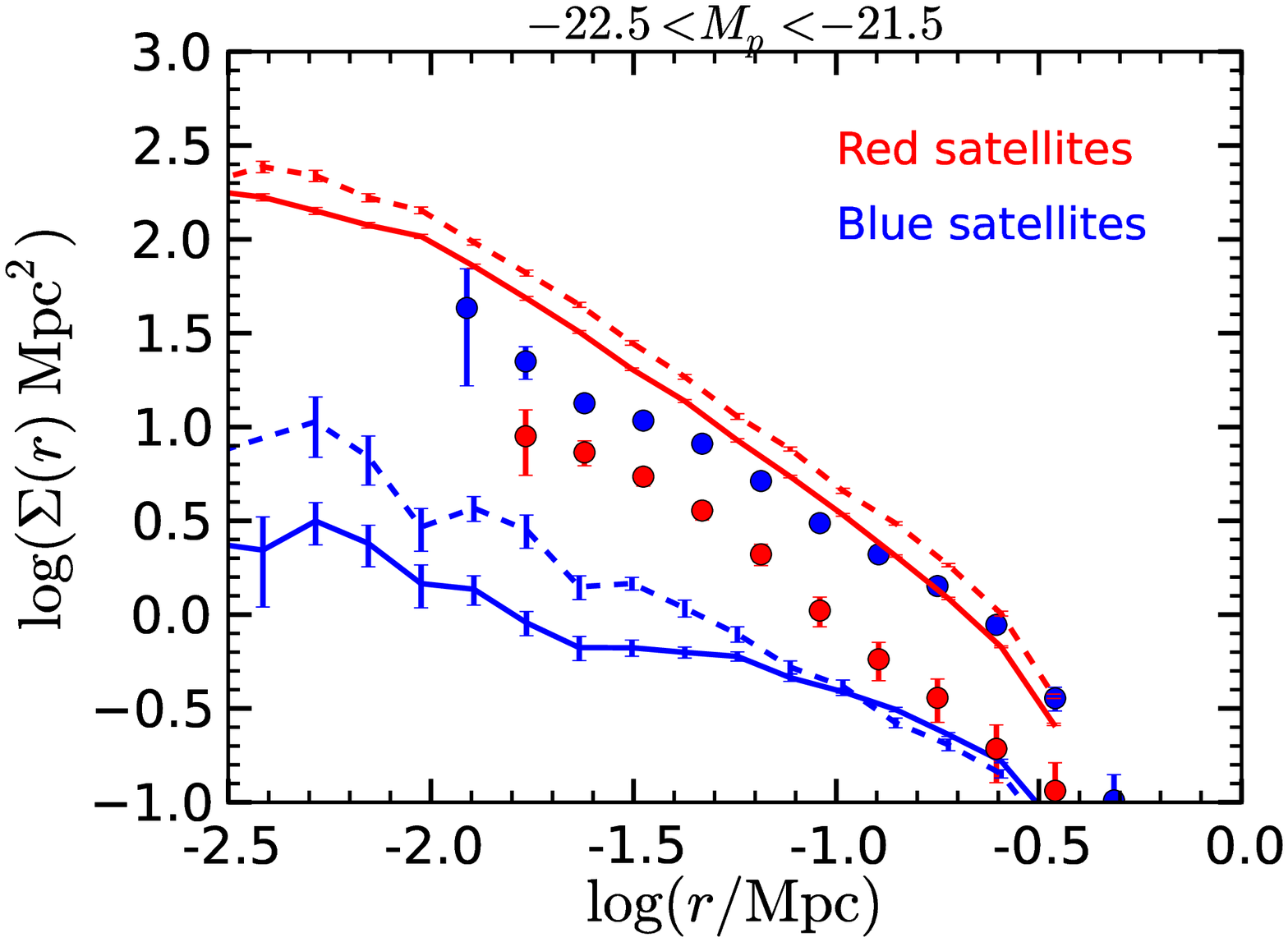}
\caption{The effect on the model satellite LF (top) and number density
    profiles (lower two panels) of changing
    the treatment of hot gas stripping in GALFORM for primary galaxies of
    magnitude $-22.5<M_p<-21.5$.  The points are the results measured from
    SDSS data, whereas solid and dashed lines are the results measured from
    catalogues constructed using the Bower et al. (2006) and Font et al.
    (2008) \gf models respectively. The middle panel shows the
    profiles of all satellites, whereas the lower panel shows them
    split by colour.}
\label{fig:font}
\end{figure}

\cite{wei12} suggest that galaxy
formation models have a generic problem with maintaining sufficient gas to form
stars in lower mass galaxies at low redshift. The satellite galaxies we
are considering are somewhat larger than those studied by
\cite{wei12}, and our choice of different colour cuts for defining
red and blue galaxies in the model and SDSS samples reduces global
systematic colour differences. For instance, the
overall fraction of blue galaxies in the model, at the magnitudes of the
satellites that we focus on, is very similar to that in the SDSS.
Satellite galaxies in our study constitute only a small fraction of
the total population of galaxies. Thus, a dramatic change in the
satellite properties will leave very little imprint on the global LF.

Alternatively, it could be that semi-analytical galaxies turn red too rapidly
after accretion into larger haloes. This idea was investigated by \cite{font08},
who changed the \gf treatment of gas stripping from subhaloes as they enter the
virial radius of large primary galaxies. Rather than hot gas being stripped from
subhaloes immediately as they enter the virial radius, a more gradual loss of
gas is adopted in the \cite{font08} model.  This allows a relatively extended period of
star formation to occur and the possiblity of bluer satellites. We have performed our
analysis on a model galaxy population constructed using this particular variant
of the \cite{bow06} \gf model. While the typical colours of galaxies do become
slightly bluer, the number of blue satellites per primary in the \cite{font08}
model increases only slightly, as can be seen in Fig.~\ref{fig:font}. The shape
of the blue satellite profile improves significantly, with the extra blue
satellites being found preferentially at small radii. However, the abundance of
red
satellites is also increased by this modification to the \gf model,
because satellites generally become more luminous as a result of the
more extended period of star formation. As a result, the \cite{font08}
model overproduces satellite abundances overall, as shown in the upper
two panels of Fig.~\ref{fig:font}. The overproduction is most
discrepant with the data at low luminosities.


Many important astrophysical processes combine to determine the
distribution of low luminosity galaxies. Therefore the distributions
of satellite galaxies will depend sensitively on aspects of the galaxy
formation model. Given that
the treatment of gas stripping can have the large impact shown in
Fig.~\ref{fig:font}, one is drawn to conclude that the ability of the 
default \cite{bow06} model to match the total satellite LF and
projected number density profile of the SDSS systems was far from inevitable.


\section{Conclusions}

Using model galaxy catalogues constructed using large dark matter
simulations and a semi-analytic galaxy formation model, we have tested
the accuracy of our procedures for measuring properties of the
distribution of satellite galaxies around bright, isolated primary
galaxies. We find that our local estimation of the abundance of
background galaxies yields unbiased estimates of the satellite galaxy
luminosity function and projected number density profile. The
agreement between results in the MS and MS-II galaxy catalogues in
their region of overlap shows that our results are numerically
converged and allows us to extend the dynamic range of the model
predictions.

Comparing the model predictions with those measured for satellite
systems in the SDSS, we find that the dependence of the satellite LF
is matched well for $\satmc=-21.0, -22.0, -23.0$ and $\Delta M_r<5$.
Lower luminosity satellites are increasingly underpredicted by the
model. The projected number density profile is also well reproduced at
radii greater than $\sim 30~{\rm kpc}$. At smaller radii, deviations
in the abundance by a factor of two are apparent. These differences
between model and SDSS are seen most strongly in the low luminosity
satellites, which show an excess in the SDSS relative to the
extrapolation of the power law from larger radii, which describes the
inner regions of the model satellite systems.

Splitting the sample into red and blue galaxies produces more dramatic
differences between the model and SDSS results. The model places a
factor $2-3$ fewer satellites around blue primaries than are present
around comparable SDSS primaries. However, the discrepancy between
model and SDSS is even larger when considering the colours of the
satellites. The model satellites are predominantly red, in contrast to
the blue-dominated SDSS satellite galaxy population. Furthermore, what
model blue satellites there are have a significantly more extended
distribution around their primary galaxy than is seen for either the
SDSS blue satellites or the red satellites in the model and SDSS.

The generally successful comparison of the \gf model with the SDSS
data performed here provides a non-trivial validation of the
assumptions and framework of this kind of modelling. At the same time,
the failure of the model to account for the observed colour dependence
of the satellite properties demonstrates that the model is incomplete
and that important physical processes, almost certainly related to the
rapidity with which infalling satellite galaxies turn red, are not
being faithfully modelled. Since a similar shortcoming is present in
the independent model of \cite{qi11}, this problem seems deep-rooted
and is worthy of further investigation.

\section*{Acknowledgements}

QG acknowledges a fellowship from the European Commission's Framework
Programme 7, through the Marie Curie Initial Training Network
CosmoComp (PITN-GA-2009-238356).  CSF acknowledges an ERC Advanced
Investigator grant 267291 COSMIWAY. This work was supported in part by
an STFC rolling grant to the Institute for Computational Cosmology of
Durham University.




\end{document}